\documentclass[twocolumn,preprintnumbers,showpacs,nofootinbib]{revtex4}

\usepackage{amssymb}
\usepackage{amsmath}
\usepackage{graphicx}

\newcommand{\dd}{\mathrm{d}} 
\newcommand{\Mpl}{M_\mathrm{pl}} 
\newcommand{\mpl}{m_\mathrm{pl}} 

\newcommand{\epsone}{\epsilon_1}
\newcommand{\epstwo}{\epsilon_2}
\newcommand{\rr}{\mathrm}
\newcommand{\ns}{n_{\rr s} }
\newcommand{\As}{A_{\rr s} }

\newcommand{\phic}{\phi_{\rr c}}

\newcommand{\phiend}{\phi_{\rr {end}}}

\newcommand{\fNL}{f_{\mathrm{NL}}}

\newcommand{\Mpc}{\rr{Mpc}}

\newcommand{\wreh}{\bar w_{\rr{reh}}}

\newcommand{\rhoend}{\rho_{\rr{end}}}

\newcommand{\be}{\begin{equation}}
\newcommand{\ee}{\end{equation}}
\newcommand{\ba}{\begin{align}}
\newcommand{\ea}{\end{align}}

\newcommand{\COSMOMC}{\texttt{COSMOMC }}

\newcommand{\squeezeup}{\vspace{-2.5cm}}

\begin{document}


\title{Updated Constraints on Large Field Hybrid Inflation}




\author{S\'ebastien Clesse} 
\email{sebastien.clesse@unamur.be}
\affiliation{Namur Center of Complex Systems (naXys), Department of Mathematics, University of Namur, Rempart de la Vierge 8, 5000 Namur, Belgium}
\author{J\'er\'emy Rekier} \email{jrek@math.fundp.ac.be} 
\affiliation{Namur Center of Complex Systems (naXys), Department of Mathematics, University of Namur, Rempart de la Vierge 8, 5000 Namur, Belgium}


\date{\today}

\begin{abstract}

We revisit the status of hybrid inflation in the light of Planck and recent BICEP2 results, taking care of  possible transient violations of the slow-roll conditions as the field passes from the large field to the vacuum dominated phase.  The usual regime where observable scales exit the Hubble radius in the vacuum dominated phase predicts a blue scalar spectrum, which is ruled out.  But whereas assuming slow-roll one expects this regime to be generic, by solving the exact dynamics we identify the parameter space for which the small field phase is naturally avoided due to slow-roll violations at the end of the large field phase.   When the number of e-folds generated at small field is negligible, the model predictions are degenerated with those of a quadratic potential.   There exists also a transitory case for which the small field phase is sufficiently long to affect importantly the observable predictions.  Interestingly, in this case the spectral index and the tensor to scalar ratio agree respectively with the best fit of Planck and BICEP2.  This results in a  $\Delta \chi^2 \simeq 5.0 $ in favor of hybrid inflation for Planck+BICEP2 ($\Delta \chi^2 \simeq 0.9$ for Planck only).  The last considered regime is when the critical point at which inflation ends is located in the large field phase.  It is constrained to be lower than about ten times the reduced Planck mass. 
The analysis has been conducted with the use of Markov-Chain-Monte-Carlo bayesian method, in a reheating consistent way, and we present the posterior probability distributions for all the model parameters.

\end{abstract}


\maketitle

\section{Introduction}

The inflationary paradigm provides an explanation to the horizon, flatness and monopole problems of the standard hot Big-Bang cosmological scenario, as well as a mechanism to generate Gaussian and nearly scale-invariant density perturbations from quantum fluctuations of one (or more than one) scalar field(s) during inflation.  Besides theoretical motivations, strong observational evidences are consistent with a primordial phase of quasi exponentially accelerated expansion.  The amplitude $\As $ and the spectral index $\ns $ of the power spectrum of primordial density perturbations have been measured with accuracy by experiments probing the Cosmic Microwave Background (CMB) temperature anisotropies, such as the Planck spacecraft~\cite{Ade:2013lta,Planck:2013kta}, the Atacama Cosmology Telescope~\cite{Das:2013zf} and the South Pole Telescope~\cite{Story:2012wx}, giving $\As = 2.196^{+0.051}_{-0.06} \times 10^{-9} $ and $\ns  = 0.9603 \pm 0.0073$ in agreement with many inflation models~\cite{Martin:2013nzq}.   A strong bound have also been established on the level of local primordial non-Gaussianities, $\fNL^{\rr{loc}} = 2.7 \pm 5.8  $ \cite{Ade:2013ydc}.   Very recently the B-mode polarization of the CMB on large scales has been measured by BICEP2~\cite{Ade:2014xna}.  The signal might be attributed to the gravitational waves produced during inflation\footnote{Note that galactic dust could contribute more importantly to the signal than initially expected, and therefore future observations will be required to affirm the discovery of primordial gravitational waves~\cite{Mortonson:2014bja,Flauger:2014qra,Ade:2014xna} }, with a tensor to scalar ratio $r = 0.20^{+0.07}_{-0.05}$ that favors super-planckian excursions of the inflaton field and points towards an energy scale associated to inflation close to the Grand-Unified energy.

In more than twenty years, hundreds of inflationary models and regimes have been proposed (for a recent review of single-field models, see~\cite{Martin:2013tda}).   
Among them the class of hybrid models is particularly interesting because they can be embedded in various high energy frameworks like supersymmetry~\cite{Dvali:1994ms,Binetruy:1996xj,Clauwens:2007wc,Lazarides:1995vr,Kallosh:2003ux,Jeannerot:2002wt,Lazarides:2007fh} and supergravity~\cite{Halyo:1996pp,Binetruy:2004hh}, Grand-Unified-Theory (GUT) ~\cite{Jeannerot:2000sv,Jeannerot:1997is,Jeannerot:2003qv,Rocher:2004et}, extra-dimensions~\cite{Fukuyama:2008dv,Fairbairn:2003yx} and string theory~\cite{Dvali:1998pa,Koyama:2003yc,Berkooz:2004yc,Davis:2008sa,Brax:2006yq,Kachru:2003sx}.  The common characteristics of hybrid models is that the field potential owns a nearly flat valley along which inflation can occur and that inflation ends with a spontaneous symmetry breaking when the field potential develops a tachyonic instability in the direction of an extra auxiliary field.  During the so-called final waterfall phase the classical field trajectories evolve towards one of the global minima of the potential, whereas in a realistic scenario a phase of tachyonic preheating is triggered~\cite{Felder:2000hj,Felder:2001kt,Copeland:2002ku} when the tachyonic mass becomes larger than the Hubble rate.

Usually the waterfall phase is assumed to be nearly instantaneous (lasting less than about one e-fold of expansion), but there exists also a generic mild waterfall regime lasting for more than 60 e-folds~\cite{Clesse:2010iz}.  In this case the observable perturbation modes exit the Hubble radius during the waterfall, changing the observable predictions of the model~\cite{Clesse:2013jra,Clesse:2012dw,Abolhasani:2010kn,Mulryne:2011ni,Kodama:2011vs,Clesse:2010iz}, and topological defects that are formed at the critical instability point of the potential can be conveniently stretched outside the observable Universe by the subsequent phase of inflation.

In the original version of the hybrid model~\cite{Linde:1993cn,Copeland:1994vg}, observable perturbation modes exit the Hubble horizon during the false-vacuum dominated phase at small field values, which is very efficient to generate many e-folds of expansion.  This is translated in the primordial scalar power spectrum by a slightly blue tilt, ruled out by Planck at more than $5 \sigma$~\cite{Ade:2013lta,Planck:2013kta}.  In its most well-known supersymmetric realizations, the F-term and D-term hybrid models~\cite{Dvali:1994ms,Binetruy:1996xj}, the scalar spectral index takes values between $0.98 \lesssim \ns \lesssim 1 $~\cite{Garbrecht:2006az,Battye:2010hg,Battye:2006pk,Pallis:2013dxa} which are disfavored by CMB observations, apart in a tuned region of the parameter space when a soft SUSY-breaking term is included to the field potential~\cite{Pallis:2013dxa}.  

The status of hybrid models became even worse with the detection of B-mode polarization by BICEP2.  If the signal is attributed to gravitational waves from inflation, it implies super-planckian field excursions, which is banned in many hybrid scenarios because supergravity corrections spoil the flatness of the potential at field values larger than the reduced Planck mass.  Nevertheless in a non-supersymmetric scenario the quantum gravity corrections can be controlled as long as the energy density and the mass remains sub-planckian~\cite{Linde:2005ht}.   Another alternative is that the potential is protected by gauge symmetries (see e.g. gauge inflation models~\cite{ArkaniHamed:2003mz,ArkaniHamed:2003wu,Kaplan:2003aj}).   

The large field regime of the original hybrid model is rather less unexplored than small field hybrid models, mostly because it is often thought that the efficient final vacuum dominated phase cannot be avoided.  This is without considering possible slow-roll violations at the transition between the two regimes.   They can prevent field trajectories to reach the slow-roll attractor at small field values if a simple condition on one of the potential parameters is satisfied~\cite{Clesse:2008pf}.  In this case the predictions of the original hybrid potential are expected to be similar to the power-law large field model.   An other potentially interesting situation is when the critical instability point is located at super-planckian field values.  

 
In this paper, we consider the original hybrid potential in the regime of large field values.  We integrate the exact field and expansion dynamics, without assuming slow-roll at the background level, and determine the model observational predictions on the scalar and tensor power spectra in each regime.  Then we use a Markov-Chain-Monte-Carlo (MCMC) method to explore the model parameter space and derive posterior probability distributions. For the first time we obtain reheating consistant cosmological constraints on potential parameters based on the Planck and BICEP2 data.   We give a particular attention to the transitory regime, for which we predict a spectral index and a tensor to scalar ratio respectively close to best fits of Planck and BICEP2.  Finally, we compare the best fit of the hybrid model to the one of the usual quadratic potential.   Note that our analysis differs from the one of Ref.~\cite{Kobayashi:2014rla} where an additional phase of inflation is added to modify the observational predictions.  It is similar but goes beyond the work of Ref.~\cite{Antusch:2014saa} since the effect of slow-roll violations is fully considered.


The paper is organized as follows:  in Sec.~\ref{sec:SFdyn} the single-field dynamics as well as the model observational predictions at first order in slow-roll expansion are introduced.  The hybrid model is presented in Sec.~\ref{sec:model}, taking care to identify all the possible regimes at large field values.  In Sec.~\ref{sec:constraints} we determine reheating consistent model constraints based on Planck and BIEP2 data, with the use of a MCMC analysis.  We conclude and discuss the perspectives of our results in Sec.~\ref{sec:ccl}.

\section{Single-Field Dynamics} \label{sec:SFdyn}

\subsection{Background}
Assuming that the Universe was filled by a homogeneous scalar field $\phi$, the Friedman-Lemaitre and Klein-Gordon equations describe the expansion and scalar field dynamics,
\begin{align} \label{eq:FL1}
H^2 & =  \frac{1}{3 \Mpl^2} \left[ \frac{\dot \phi^2}{2} + V(\phi)  \right]~, \\
\label{eq:FL2}  \frac{\ddot a}{a } & =  \frac{1}{3 \Mpl^2} \left[ -\dot \phi^2 + V(\phi)   \right]~, \\
\label{eq:KG}   &  \ddot \phi + 3 H \dot \phi + \frac{\partial V}{\partial \phi} = 0~,
\end{align}
where $a$ is the scale factor, $H$  the Hubble rate, $\Mpl \equiv \mpl / \sqrt{8 \pi} $ the reduced Planck mass, $V(\phi)$ is the scalar field potential and where a dot denotes the derivative with respect to the cosmic time $t$.  
The slow-roll approximation consists in neglecting the kinetic terms in Eqs.~(\ref{eq:FL1}) and (\ref{eq:FL2}) as well as the second time derivatives of the field in Eq.~(\ref{eq:KG}).   Since we are interested by transient slow-roll violations the exact dynamics have been integrated numerically.   It is compared to the slow-roll approximation in Sec.~\ref{sec:model}.   
Using the number of e-fold $N \equiv \ln a/a_{\rr i} $ as the time variable, those equations can be rewritten as 
\begin{align}
H^2&=\frac{V(\phi)}{3-\frac{1}{2}\left(\frac{d\phi}{dN}\right)^2},\\
\frac{1}{H}\frac{dH}{dN}&=-\frac{1}{2}\left(\frac{d\phi}{dN}\right)^2,\\
\frac{1}{3 -\frac{1}{2}\left(\frac{d\phi}{dN}\right)^2}\frac{d^2\phi}{dN^2}& +\frac{d\phi}{dN}=-\frac{d\ln V}{d\phi}.
\end{align}
In this form, the field dynamics does not depend on the Hubble rate.  It is then usual to introduce the Hubble flow functions, also referred as slow-roll parameters, 
\begin{align}
\epsone & \equiv - \frac{\dot H}{H^2} = \frac{1}{2}\left(\frac{d\phi}{dN}\right)^2 \simeq \frac{\Mpl^2}{2} \left( \frac{V_{,\phi}}{V} \right)^2~,  \\
\epstwo & \equiv \frac{\dd  \ln \epsone}{\dd N} \simeq 2 \Mpl^2 \left[  \left( \frac{V_{,\phi}}{V} \right)^ 2 - \frac{V_{,\phi \phi}}{V}  \right] ~,\\
\epsilon_{i>2} & \equiv \frac{\dd  \ln |\epsilon_{i-1}|}{\dd N},
\end{align}
where the approximate expressions are obtained under the slow-roll approximation, valid as long as $\epsone $ and $\epstwo$ are much smaller than one. 

\subsection{Linear Perturbations}

Measuring the temperature anisotropies and the B-mode polarization of the CMB gives access to the statistical properties of the primordial curvature perturbations $\zeta$ and tensor perturbations $h$.  These properties are encoded in the $n$-point correlation functions.  The two-point correlation function is the integral of the adimensional power spectrum $\mathcal P(k)$  over the logarithm of the wavenumbers.  
By solving the perturbed Einstein equations at second order in terms of slow-roll parameters, and assuming the initial states to be the Bunch-Davis vacuum,  analytical expressions for the scalar and tensor perturbation power spectra can be derived~\cite{Gong:2001he}.   Expanding these spectra around a chosen pivot scale $k_* $ (usually $k_* = 0.05\ \Mpc^{-1} $), one gets for scalar perturbations 
 \be
 \mathcal P_{\zeta,h} (k) = \mathcal P_{\zeta_0,h_0} \times \left[ a_0 + a_1 \ln \left( \frac{k}{k_*} \right)  \right]  ~, 
 \ee
 with
 \be \label{eq:P0}
 \mathcal{P}_{\zeta_0}=\frac{H_*^2}{8\pi^2 \Mpl^2 \epsilon_{1*}} ~,
 \ee
 and where the star subscript denotes quantities evaluated at the time $t_*$ when the scale $k_*$ exits the Hubble horizon, $k_* = a(t_*) H(t_*)$.  To first order, the coefficients of the expansion read 
\begin{align}
a_0^{(\rr s)}&=1-2(C+1)\epsilon_{1*}-C\epsilon_{2*}+ \mathcal O(\epsilon^2)\\
a_1^{(\rr s)}&=-2\epsilon_{1*}-\epsilon_{2*}+\mathcal O(\epsilon^2)\\
a_0^{(\rr t)}&=1-2(C+1)\epsilon_{1*}+\mathcal O(\epsilon^2)\\
a_1^{(\rr t)}&=-2\epsilon_{1*}+\mathcal O(\epsilon^2),
\end{align}
with $C \equiv \gamma_\text{E}+\ln2-2$, $\gamma_\text{E}$ being the Euler constant.  In this notation, $\As$ can thus be identified to $\mathcal P_0 a_0^{(\rr s)}$. 

At leading order, the power spectrum of tensor perturbations is given by
 \be
 \mathcal{P}_{h_0}=\frac{2H_*^2}{\pi^2 \Mpl^2}.
\ee
 which gives a tensor to scalar ratio 
 \be \label{eq:running}
r\equiv \frac{\mathcal{P}_{h_0}}{\mathcal{P}_{\zeta_0}}=16\epsilon_1^*, 
 \ee
The scalar spectral index is defined as 
\be
\ns -1\equiv \frac{\dd \ln \mathcal P_\zeta(k) }{\dd \ln k}.
\ee
 At first order in slow-roll parameters, this gives
 \be \label{eq:ns}
 \ns = 1 - 2 \epsilon_{1*} - \epsilon_{2*},
 \ee
 and therefore one can relate the shape of the scalar and tensor power spectra to the background dynamics at the time when the pivot scale exits the Hubble horizon, i.e. about $N_* \sim 60$ e-folds before the end of of inflation.
 In Sec.~\ref{sec:model}, those relations are applied to the hybrid potential to derive the model observable predictions.  
 

\subsection{Reheating}

The physical size of the pivot mode as it crosses the horizon can be written as 
\begin{equation}
\frac{k_*}{a_*}=\frac{k_*}{a_0}\frac{a_0}{a_\text{end}}\frac{a_\text{end}}{a_*}.
\end{equation}
$k_* /a_0$ is the physical size of the pivot scale now. The evolution after inflation is contained in $a_0 /a_\text{end}$. The whole relation can be conveniently parametrised by
\begin{equation}
\frac{k_*}{a_*}=\frac{k_*}{a_0}\left(\frac{\rho_\text{end}}{\rho_{\gamma 0}}\right)^{1/4}R_\text{rad}^{-1}\frac{a_\text{end}}{a_*}.
\end{equation}
The new parameter $R_\text{rad}$ is equal to $1$ in the case of instantaneous reheating after inflation.  Otherwise it is related to the mean equation of state parameter $\wreh$ during the reheating era and to the reheating energy $\rho_{\rr{reh}}$ through~\cite{2010PhRvD..82b3511M}
\be
\ln R_{\rr{rad}} = \frac{1 - 3 \wreh}{12(1+\wreh)}\ln \left( \frac{\rho_{\rr{reh}}}{\rho_{\rr{end}}} \right).
\ee 
The reheating parameter plays an important role in fixing $N_*$, the number of e-fold realized between $t_*$ and the end of inflation.  It is then convenient to introduce the reheating parameter $R \equiv R_{\rr {rad}} \rhoend^{1/4} /\Mpl $, so that one has
\be
\frac{k_*}{a_*}=\frac{k_*}{a_0} \left(\frac{\Mpl}{\rho_{\gamma 0}^{1/4} }\right) \rhoend^{1/2} R^{-1} \rr e^{N_*},
\ee
which makes  $N_*$ invariant under a rescaling of the scalar field potential.   In Sec.~\ref{sec:constraints} the reheating $R_{\rr{rad}}$ parameter has been included within the Markov-Chain-Monte-Carlo analysis in order to derive reheating consistant constraints on the large field hybrid model. 

\section{Hybrid Model}  \label{sec:model}

\subsection{Field Potential}

The original two-field hybrid potential reads~\cite{Linde:1993cn,Copeland:1994vg}
\be
V(\phi,\psi) = \Lambda^4 \left[ \left( 1 - \frac{\psi^2}{M^2} \right)^2 + \frac{\phi^2}{\mu^2} + \frac{2 \phi^2 \psi^2}{\phic^2 M^2}  \right]~.
\ee
It owns a nearly flat valley in the direction $\psi = 0$ along which inflation occurs, with the effective single-field potential 
\begin{equation}
V(\phi)=\Lambda^4\left(1+\frac{\phi^2}{\mu^2}\right).
\end{equation}
When the inflaton reaches the critical value $\phic $, the potential develops a tachyonic instability forcing the fields to reach one of the global minima of the potential, at $(\phi, \psi) = (0,\pm M)$.  In the following, we do not assume a specific high-energy framework and consider the possibility to have inflation at field values larger than the Planck scale.  

The inflationary valley can be reached from field values exterior to it without any important fine-tuning, as shown in Refs~\cite{Clesse:2008pf,Clesse:2009ur,Clesse:2009zd,Easther:2013bga,Easther:2014zga}.   Along the valley, the dynamics can be decomposed in two phases: 
i) at large field values $\phi > \mu$, when the potential is of quadratic form, and ii) at small field values $\phi < \mu$ when the potential is dominated by the false vacuum term.  At the end of inflation, the waterfall phase takes place and we assume it to be nearly instantaneous throughout the paper, except in Sec.~\ref{sec:mild waterfall} where the case of a mild waterfall at large field values is discussed briefly.  

Finally the parameter $\Lambda$ fixes the energy scale of inflation.  It only influences the e-fold time $N_*$ and has no impact on the background field dynamics.  That makes a 4-dim parameter space, to which will be added standard cosmological parameters and nuisance parameters in Sec.~\ref{sec:constraints}.

Along the valley, the Hubble-flow parameters in the slow-roll approximation are given by 
\begin{subequations}
\label{eps:HIV}
\begin{align}
\epsilon_1^{\rr{SR}} =& \frac{2 \Mpl^2 \frac{\phi^2}{\mu^2} }{\mu^2 \left( 1 + \frac{\phi^2}{\mu^2} \right)^2}~,\\
\epsilon_2^{\rr{SR}} =&  \frac{4 \Mpl^2 \left( -1 + \frac{\phi^2}{\mu^2} \right) }{\mu^2 \left( 1 + \frac{\phi^2}{\mu^2} \right)^2}~.
\end{align}
\end{subequations}
They are represented on Fig. \ref{fig:eps1vsphi} and Fig.~\ref{fig:ploteps2} for different values of the $\mu$ parameter.  
 In the vacuum dominated regime, inflation stops at the critical point $\phi_c$ below which the potential develops the tachyonic instability.  Since $\epsilon_1 \ll 1$ and $\epsilon_2 < 0$, a blue spectrum of scalar perturbation is expected.     In the large field regime, one gets $\epsilon_1 \simeq 2 \Mpl^2 / \phi^2 $ and $ \epstwo \simeq 4 \Mpl^2 / \phi^2 $ as for the massive potential, and thus the spectral index is red.  However, assuming the slow-roll dynamics is valid, the vacuum dominated phase is so efficient in terms of e-folds generation ($N \gg 60$ generically) that  observable scales necesseraly exit the horizon during this phase.  But in the case where $\mu \lesssim 1.6 \Mpl$, the slow-roll conditions are not satisfied during the transition between the large field and the vacuum dominated phase, and it has been shown that in such a case the kinetic energy acquired by the the inflaton field prevents inflation from taking place in the small field regime~\cite{Clesse:2008pf}.  Another possibility is that the critical point is located at large field values.  In both cases, Hubble exit of observable scales occurs in the large field phase, which generates a red spectrum possibly in agreement with observations.   Those regimes are studied in details in the next section.

\subsection{Large Field Regimes}

\subsubsection{Chaotic-like: $\phic < \mu  \ll \phi_*$ and $\mu < \Mpl $  }

The first considered regime is the one similar to chaotic inflation with a quadratic potential.   Inflation terminates at the end of the large field phase and is not triggered back afterwards.  
It is important to emphasize that this possibility exists only due to the effect of slow-roll violations during and after the transitory phase between large field and small field values:  when the slow-roll is strongly violated at the transition close to $\phi = \mu$, trajectories gain sufficient velocities to prevent them to reach back the slow-roll attractor at small field values~\cite{Clesse:2008pf}.  This effect is not trivial since the slow-roll dynamics is violated at small field values whereas the slow-roll conditions are apparently satisfied ($\epsone^{\rr{SR}} \ll 1 $ and $\epstwo^{\rr{SR}} \ll 1$) and one has to integrate for the exact dynamics to put this in evidence.   Assuming slow-roll at small field values, one would obtain that the large field phase lies outside the range of observable modes.  But in reality, instead of stopping at the critical instability point after an efficient small field phase, inflation stops when the first Hubble flow parameter $\epsone$ reaches unity at the end of the large field phase. 

\begin{figure}[h]
  \begin{center}
    \resizebox{0.5\textwidth}{!}{\input{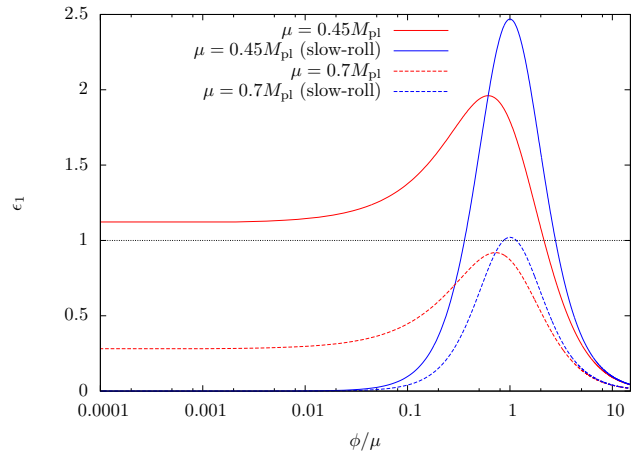}}
  \end{center}
  \caption{Evolution of the first Hubble-flow parameter $\epsilon_1$ as a function of $\phi / \mu$, computed by solving the full dynamics or using the slow-roll approximation.  The full dynamics solution differs greatly from the slow-roll solution at small values of $\mu$.  The kinetic energy acquired at the transition between the large field and the vacuum dominated phase prevents inflation to take place at small field values.    
   \label{fig:eps1vsphi}
 }
\end{figure}

\begin{figure}[h]
  \begin{center}
    \resizebox{0.5\textwidth}{!}{\input{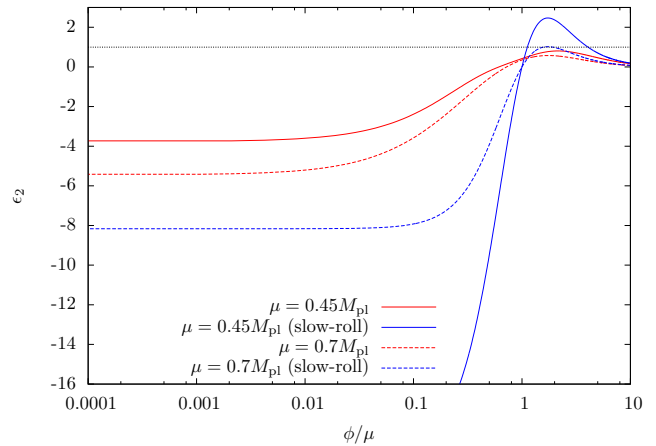}}
  \end{center}
  \caption{Evolution of the second Hubble-flow parameter $\epsilon_2$ as a function of $\phi / \mu$, computed by solving the full dynamics or using the slow-roll approximation.  As in Fig.~\ref{fig:eps1vsphi} the full dynamics differs  from the slow-roll approximation for small values of $\mu$, leading to different observable predictions.  
   \label{fig:ploteps2}
}
\end{figure}

\begin{figure}[h]
  \begin{center}
    \resizebox{0.5\textwidth}{!}{\input{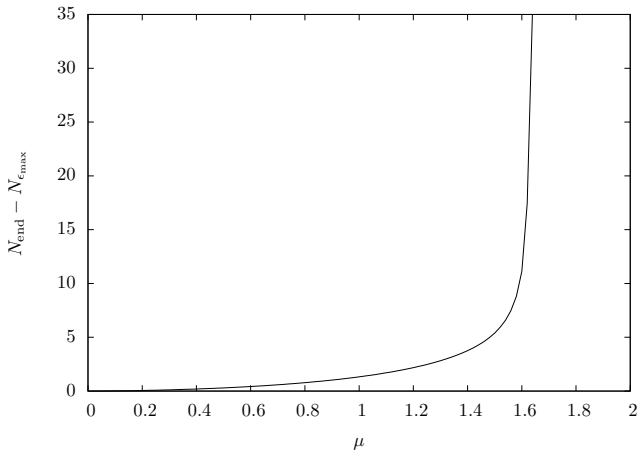}}
  \end{center}
  \caption{Number of e-fold produced after reaching the maximum of $\epsilon_1$ as a function of $\mu$.  Below the threshold value $\mu_{\rr{thr}}\sim 1.6 \Mpl$, only a few e-folds are realized at small field values in contradiction with slow-roll predictions.}
  \label{fig:DNvsmu}
\end{figure}

Fig.~\ref{fig:eps1vsphi} shows the value of  $\epsilon_1$ as a function of $\phi$, for two representative values of $\mu$ ($\mu=0.4 M_\text{pl}$ and $\mu=0.7 M_\text{pl}$) both with and without using the slow-roll approximation.  When $\mu$ is smaller than some threshold value, $\epsilon_1$ does not decrease below one at small field and inflation is not triggered again, contrarily to what is expected in the slow-roll approximation with $\epsone$ decreasing down to tiny values.  Fig.~\ref{fig:ploteps2} shows the evolution of $\epsilon_2$ in a similar fashion.  Here again we find an important difference between slow-roll value and exact values.   The influence of $\mu$ on this effect is shown on Fig. \ref{fig:DNvsmu} with the number of e-fold of expansion after the maximum of $\epsilon_1$ plotted against $\mu$.  Below $\mu=\mu_{\text{thr}}\sim 1.6 \Mpl$, only a reduced number of e-folds are realized, and for $\mu \lesssim 0.8 \Mpl$ it is marginal (lower than unity).  In the latter case, the small field phase does not affect significantly the observable predictions, which corresponds to the chaotic-like regime of hybrid inflation.  

For the derivation of the observable predictions, the slow-roll approximation can be used up to the point where $\epsone = 1$, corresponding to a final field value
\be
\phiend \simeq \frac{\sqrt 2}{2} \Mpl \left( 1+ \sqrt{1 - \frac{\sqrt 2 \mu^2}{\Mpl^2}} \right)~.
\ee
In the limit $\mu \ll \Mpl$, one recovers the expected value for a quadratic potential $\phiend \simeq \sqrt 2 \Mpl $.   
Note that the exact value can differ significantly from this slow-roll value, but this only has an non-significant effect on the value of $\phi_*$.
By integrating the slow-roll equation
\begin{equation}
\frac{\dd\phi}{\dd N}=- \Mpl^2\frac{\dd \ln V}{\dd\phi},
\end{equation}
one obtains 
\begin{equation} \label{eq:HI_SR_dyn}
\frac{\mu^2}{2M_\text{pl}^2}\left[ \ln \frac{\phiend}{\phi_*} + \frac{1}{2\mu^2}(\phiend^2-\phi_*^2)\right]=N(\phi_*)-N_\text{end}.
\end{equation}
which can be inverted to get the field value $\phi_*$ at the time of Hubble crossing of the pivot scale $k_*$.  
It is then straightforward to calculate the scalar power spectrum amplitude and spectral index as well as the tensor to scalar ratio with the use of Eqs.~(\ref{eq:ns}) and (\ref{eq:running}).
In the limit $\mu \ll \Mpl$, one finds 
\begin{align}
\phi_* & \simeq 2 \Mpl \sqrt{N_* + \frac 1 2 } \simeq 15.5 \Mpl ~,\\
\epsilon_{1*} & \simeq \frac{1}{2 N_* + 1} \simeq 0.00826 ~, \hspace{1mm} \epsilon_{2*}  \simeq 2 \epsilon_{1*} \simeq   0.0165 ~, \\
\ns & \simeq 1 - \frac{4}{2 N_* +1}  \simeq  0.967 , \hspace{1mm} r  \simeq \frac{16}{2 N_* + 1} \simeq 0.132.
\end{align}
Those values degenerated with the predictions for a quadratic potential are obtained assuming $N_* = 60$ and are in agreement with both Planck and BICEP2 data.   The chaotic-like regime corresponds to the bottom left part of Figs.~\ref{fig:contourns} and \ref{fig:contourr} where the scalar spectral index and the tensor to scalar ratio are represented in the plane $(\log_{10} \mu, \log_{10} \phic)$ using the exact background dynamics.
 
\subsubsection{Transitory: $\phic < \mu  \sim \phi_*$ and $\mu \sim \Mpl $ }

The second considered regime is the transitory case where $\mu$ is close to the threshold value $\mu_{\rr{thr}} $ below which slow-roll violations prevent the last 60 e-folds of inflation to occur in the small field phase. 
As shown on Fig.~\ref{fig:DNvsmu} between $\mathcal O(1)$ and $\mathcal O(60)$ e-fold can be realized in the vacuum dominated phase (small field values).   Nevertheless, observable scales still exit the Hubble radius during the large field phase.   It results that $\epsilon_{1*} $ and $\epsilon_{2*}$ take larger values than in the previous chaotic-like regime, depending on the duration of the vacuum dominated phase.   Therefore the scalar spectral index is lowered and can accommodate the best fit of Planck at $\ns = 0.961$.   Simultaneously the tensor to scalar ratio is enhanced and can accommodate the central value of BICEP2 $r = 0.20$.   This regime is thus favored by CMB data compared to the case of a quadratic potential, even if both models have parameter space within the $2 \sigma $ confidence level, assuming $N_* = 60$.

On Figs.~\ref{fig:contourns} and \ref{fig:contourr}  the predictions for the spectral index and the tensor to scalar ratio are displayed, using the exact background dynamics and assuming $N_* = 60$.  The figure shows how the observable predictions change when varying the parameters $\mu$ and $\phi_c$ that fixes the end of inflation.   For $\phic \ll \Mpl$, the spectral index is close to the best fit of Planck when $\mu \sim 2-3 \Mpl$, as well as in a very thin band at $\mu \simeq 4 \Mpl$.   Increasing $\phic$ up to $\phic \sim \mu$, one gets that the best fit is obtained at $3 \Mpl <\mu < 5 \Mpl $.   This is expected since the increase of $\phic $ tends to reduce the number of e-folds generated in the vacuum dominated phase, larger values of $\mu$ thus being necessary to make this phase more efficient.  

Finally, note that the transition between the transitory regime and the usual small field regime where all the relevant e-folds are realized in the vacuum dominated phase (and predicting a blue scalar power spectrum excluded by observations) is found to be very abrupt.  

\subsubsection{Large critical field value:  $\mu < \phic  < \phi_* $ }

In this third regime, the critical instability point $\phic$ below which field trajectories are destabilized is located at in the large field phase, so that the conditions $\mu  < \phic$ and $\phiend = \phic >  \sqrt 2 \Mpl$  are satisfied.  

\begin{figure}
  \begin{center}
  \includegraphics[width=0.45 \textwidth]{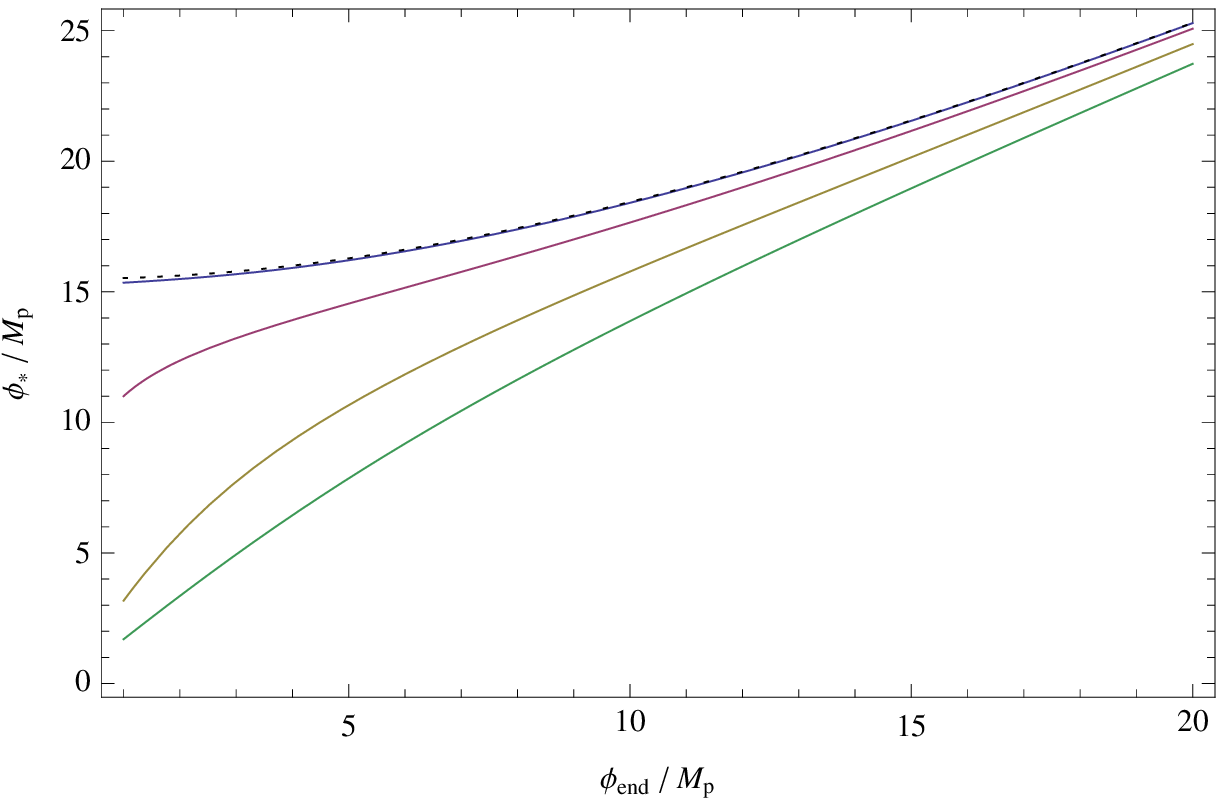}
   \includegraphics[width=0.45 \textwidth]{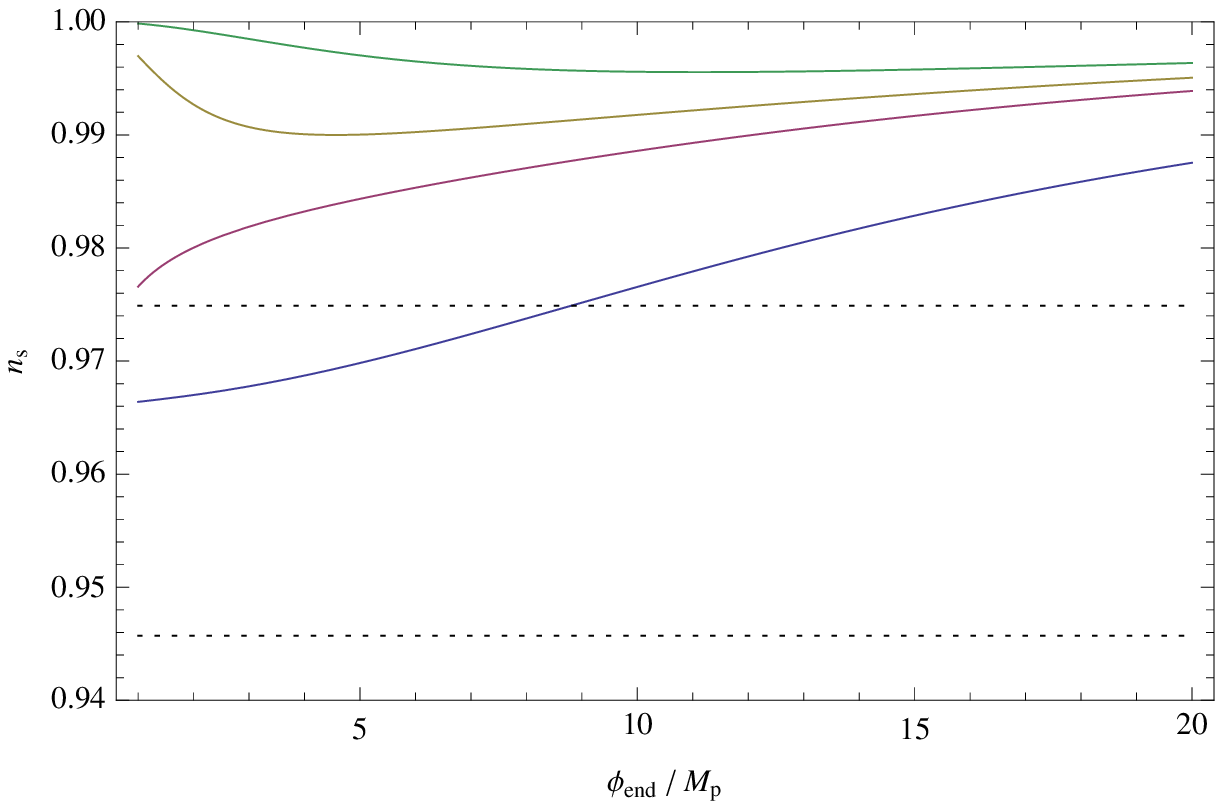}
   \includegraphics[width=0.45 \textwidth]{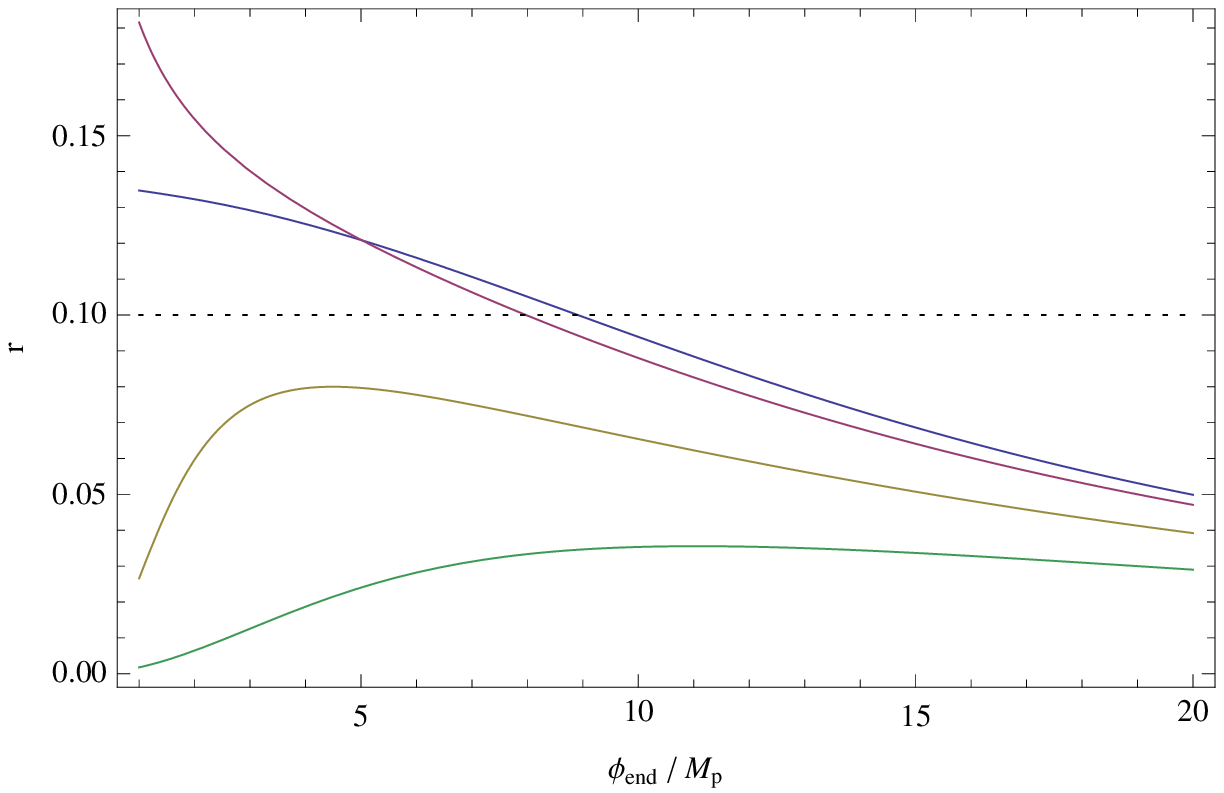}
  \end{center}
  \caption{$\phi_*$ (top) and corresponding $n_s$ (central) and $r $ (bottom) plotted as a function of $\phi_{\rr{end}} = \phic$ using Eq.~\ref{eq:phistar}, for $\mu = 1 \Mpl$ (blue), $\mu = 5 \Mpl$ (red), $\mu = 10 \Mpl$ (yellow) and $\mu = 15 \Mpl$ (green), assuming $N_* = 60$.  The horizontal dotted lines in the central and bottom panels represent respectively the $2\sigma$ regions of Planck and BICEP2.  The dotted line in the top panel is obtained by using the approximation of Eq.~\ref{eq:phistarbis}.  }
  \label{fig:largephic}
\end{figure}

\begin{figure*}
  \begin{center}
    \resizebox{0.8\textwidth}{!}{\includegraphics[angle=-90,origin=c]{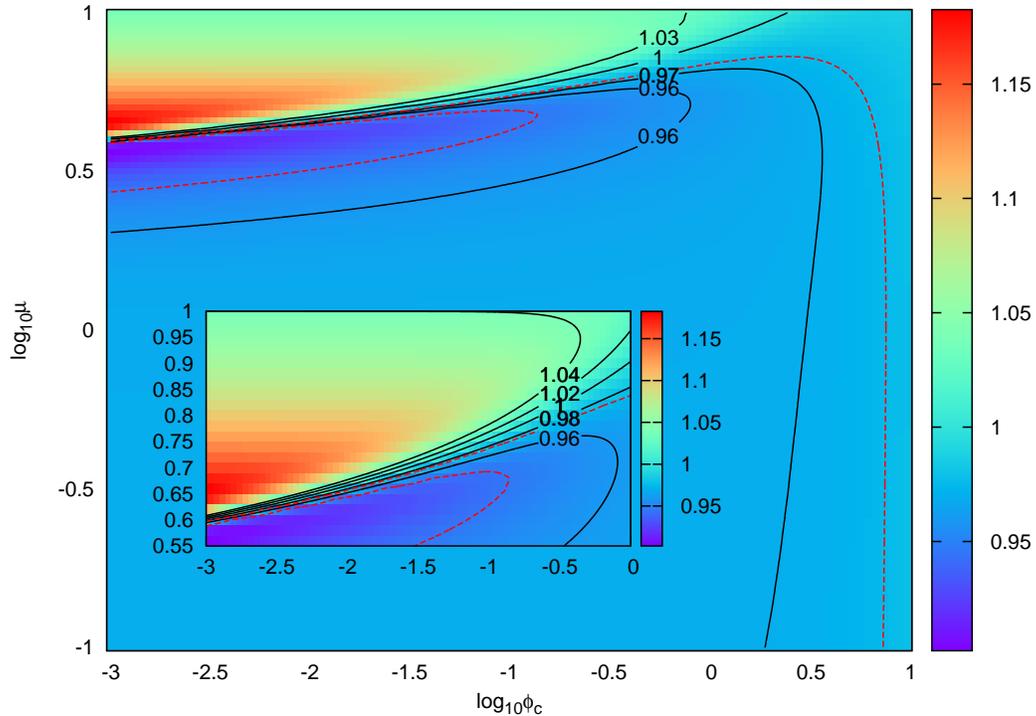}}
  \end{center}
  \squeezeup
  \caption{Contour plot of the spectral index $n_s$ in the plane ($\log_{10} \phic, \log_{10} \mu)$, using the exact background dynamics and assuming $N_* =60$.  The red dashed contours represent the $2\sigma$ confidence interval for Planck.}
  \label{fig:contourns}
\end{figure*}

\begin{figure*}
  \begin{center}
   \resizebox{0.8\textwidth}{!}{\includegraphics[angle=-90,origin=c]{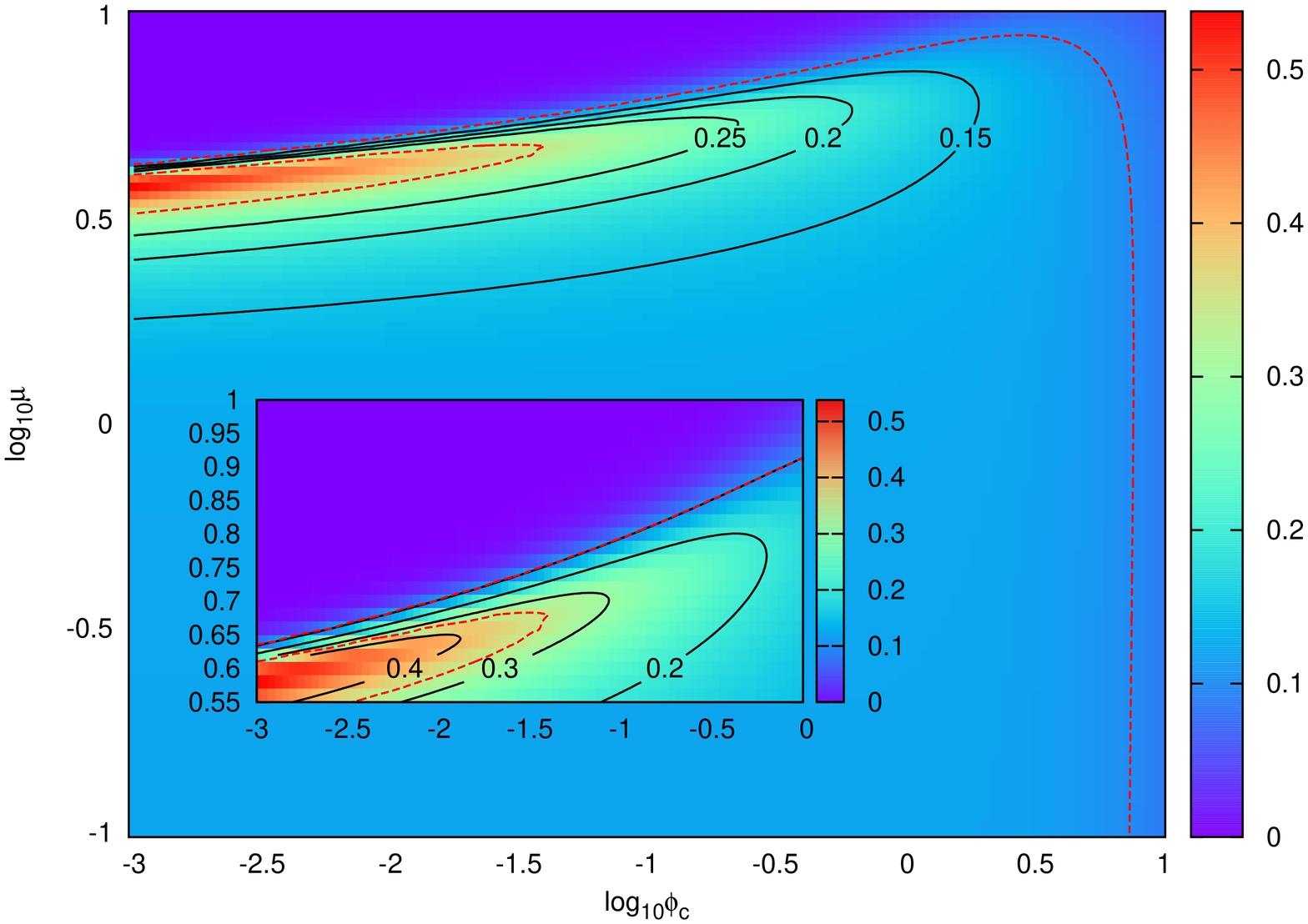}}
  \end{center}
  \squeezeup
  \caption{Contour plot of the tensor to scalar ratio $r$ in the plane ($\log_{10} \phic, \log_{10} \mu)$, using the exact background dynamics and assuming $N_* =60$.  The red dashed contours represent the $2\sigma$ confidence interval for BICEP2.}
  \label{fig:contourr}
\end{figure*}

The slow-roll approximation is valid prior to the critical point, and thus Eq.~(\ref{eq:HI_SR_dyn}) can be used to derive the corresponding observable predictions.  It can be inverted to find $\phi_* $ in terms of the the principal branch of the Lambert function $W_0(z) $, 
\be  \label{eq:phistar}
\phi_*^2 = \mu^2 W_0 \left( \frac{\phiend^2}{\mu^2} \rr e^{ \frac{\phiend^2 + 4 N_*}{\mu^2}}  \right)~.
\ee
In the limit $\mu \ll \phic $, one has 
\be \label{eq:phistarbis}
\phi_*^2 \simeq \phic^2 + 4 N_* \Mpl^2
\ee
which gives
\be
\ns  \simeq 1 - \frac{8 \Mpl^2}{  \phic^2 + 4 N_* \Mpl^2}
\ee
and
\be
r \simeq \frac{32 \Mpl^2} { \phic^2 + 4 N_* \Mpl^2}
\ee
independent of the value of the parameter $\mu$.   Values of $\phic$ larger than the planck mass therefore correspond to spectral index values closer to unity, reducing the agreement with observations.  
On Fig.~\ref{fig:largephic} we have plotted $\phi_*$, $n_s$ and $r$ as a function of $\phiend$ for several values of the parameter $\mu$ and assuming $N_* = 60$.   Note that those analytical results are in agreement with numerical ones, displayed on Figs.~\ref{fig:contourns} and~\ref{fig:contourr} (right part of the plot).  

As shown in Fig.~\ref{fig:contourns}, values of $\phic \gtrsim 10 \Mpl $ are excluded by Planck at more than $2\sigma$ level.   Therefore hybrid inflation in the regime of large critical field show important deviations compared to the observable predictions of the chaotic large field regime, and Planck data constrain the critical value that must be at maximum of the order of the Planck mass.

\subsubsection{Mild waterfall:  $\phi_* < \phic $ and $M \mu \gtrsim \Mpl^2 $  }  \label{sec:mild waterfall}

Finally, it is important to mention that a mild waterfall phase is possible after crossing the instability point.  In this case the last 60 e-folds of inflation can be realized during the waterfall, as first shown in Ref.~\cite{Clesse:2010iz}.   This possibility and the resulting modifications of the observable predictions have been studied in details in Refs.~\cite{Clesse:2013jra,Clesse:2012dw,Abolhasani:2010kn,Kodama:2011vs,Clesse:2010iz} for sub-planckian field values and in Ref.~\cite{Mulryne:2011ni} for super-planckian fields.  Below we briefly comment on the large field case.

A condition for the waterfall to be mild is that $\mu M \gtrsim \Mpl^2$.  In order to calculate the corresponding observable predictions, one has to solve the multi-field dynamics both at the background and linear perturbation levels, or to use of the $\delta N$ formalism.   The latter option has been implemented numerically in Ref.~\cite{Mulryne:2011ni} and regions where the spectral index is in agreement with observations have been found.  
In the generic case where $\psi_* \ll M$, a large level of non-gaussianity can be produced with 
\be
\fNL \approx \frac{10 \Mpl^2}{3 M^2},
\ee 
which is now ruled out by Planck if the parameter $M$ is lower than the Planck mass.   It is nevertheless possible to find parameters for which the spectral index is close to the Planck best fit and producing a negligible amount of non-Gaussianity.  We did not explore further this regime that requires the implementation of the multi-field dynamics and the $\delta N$ formalism, and we leave for a future work the complete statistical analysis of the super-planckian mild waterfall case.

\section{Model Constraints}  \label{sec:constraints}

\begin{figure*}[t]
  \begin{center}
    \resizebox{0.95\textwidth}{!}{\includegraphics[angle=0,origin=c]{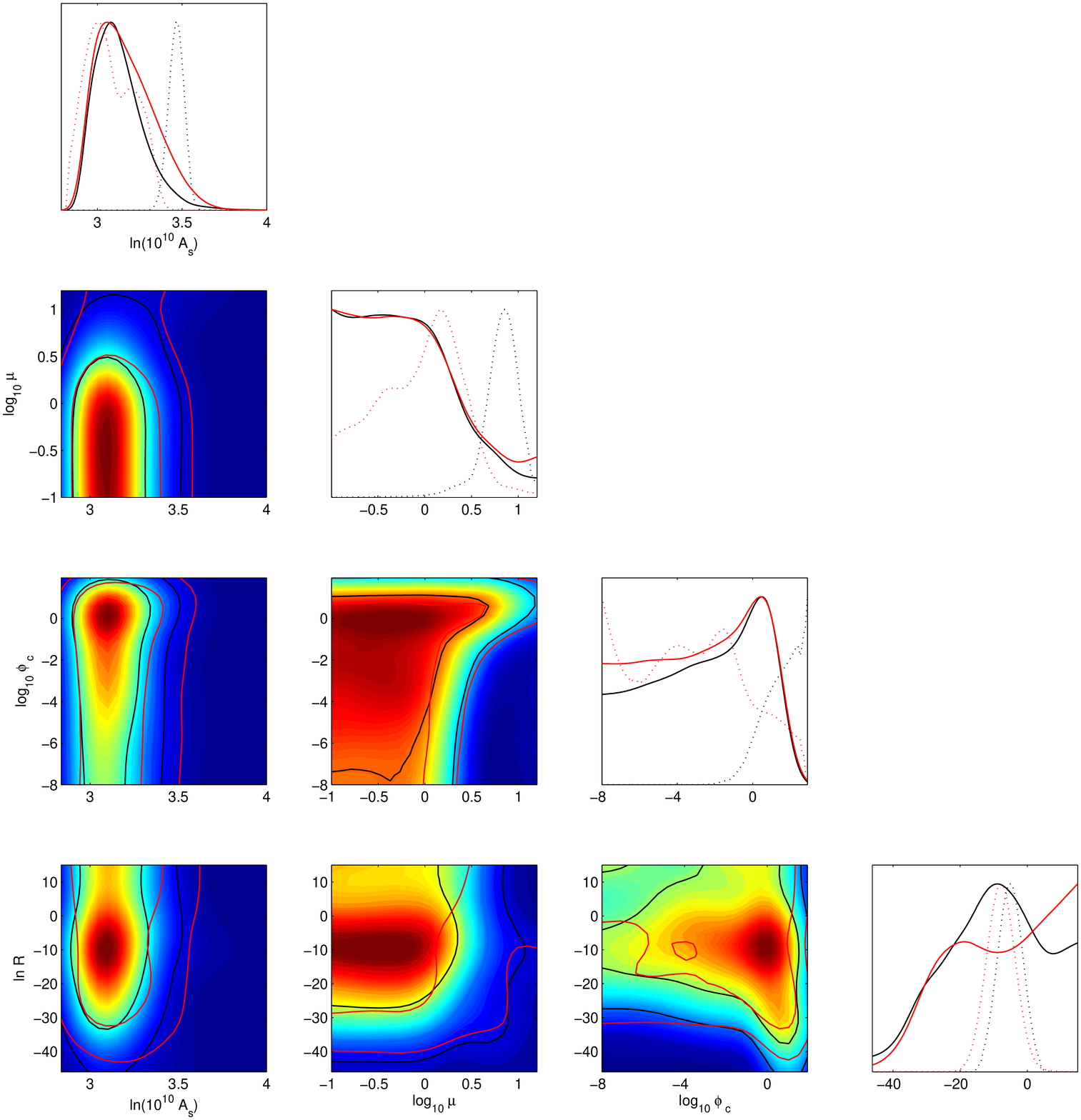}}
  \end{center} 
  \caption{Marginalized one-dimensional and two-dimensional posterior probabilities for the hybrid model parameters (in reduced Planck mass units) in the large field regime, for Planck+BICEP2.  The red contours are the $1\sigma$ and $2\sigma$ regions of confidence.  The black contours are the $1\sigma$ and $2\sigma$ regions for Planck only.  In the 1D plots, the black/red solid lines show the marginalized posterior distributions of the parameters respectively for Planck and Planck+BICEP2.  The dotted lines represent the mean likelihoods.     }
  \label{fig:MCMCtri}
\end{figure*}

In this section we derive updated constraints on the hybrid model parameter space by performing a Markov-Chain-Monte-Carlo Bayesian statistical analysis.   For this purpose we use a modified version of the \COSMOMC numerical package~\cite{Lewis:2013hha}.   In Bayesian inference the posterior probability of model parameters $\lambda_i$ given some data $D$ (assuming that the model is the true one) are given by Bayes' theorem 
\be
p(\lambda_i  | D ) = \frac { p(D| \lambda_i)  \pi(\lambda_i) }{\int \rr d \lambda_i p(D| \lambda_i ) \pi(\lambda_i)},
\ee
where $\pi(\lambda_i) $ is the prior probability distribution for the parameter $\lambda_i$, and where the denominator is a normalisation factor called the Bayesian evidence.   For the purpose of constraining model parameters the Bayesian evidence can be ommitted.   

\subsection{Priors}

The choice of the prior can play a crucial role.  In the case of the hybrid model, there is no a priori information on how small compared to the Planck scale can be the position of the critical instability point $\phic$, the false-vacuum $ \Lambda$, and the slope along the valley described by the parameter $\mu$.   The magnitude of the reheating parameter $R$ is also not a priori known.   As a consequence we have considered Jeffrey's priors on these parameters, which is an uniform prior on a logarithmic scale.   Note that an alternative choice of parameter is the scalar field mass $m = \sqrt{\Lambda} / \mu$, which remains small compared to the Planck scale.  But it is straightforward to derive the posterior probability of $ m$ (assuming a Jeffrey's prior) from $\Lambda$ and $\mu$ posteriors  by using importance sampling.  

Looking at Eq.~(\ref{eq:P0}) one sees that $\Lambda$ and $\mu$ both contribute to the scalar spectrum amplitude, which is tightly constrained by Planck.  So there is a high level of degeneracy between $\Lambda$ and $\mu$, and the sampling method will loose a lot of efficiency if it probes these two parameters.   As noticed in Ref.~\cite{Martin:2013nzq} it is more convenient to replace $\Lambda$ by the scalar spectrum amplitude with a logarithmic flat prior.  In our analysis, $\Lambda$ is thus a derived parameter together with the energy density at the end of inflation $\rhoend$.   One could also derive the reheating energy $\rho_{\rr{reh}}$ assuming a mean equation of state parameter $\bar w$.   However, in the case of hybrid inflation, reheating does not proceed with coherent field oscillations but with a phase of tachyonic preheating.  Deriving $\bar w$ therefore requires the use of lattice simulations, which is beyond the scope of this paper.

Regarding the ranges of the parameters, $\mu$ cannot take arbitrary large values because at some point quantum stochastic effects are expected to dominate over the classical dynamics~\cite{Martin:2011ib,Levasseur:2013tja}.  But for $\mu \gtrsim \Mpl$ inflation occurs in the vacuum dominated phase and the scalar power spectrum is blue, which is ruled out by observations.   As a consequence the contribution to the Bayesian evidence of this part of the parameter space will be negligible.  At the opposite, tiny values of $\mu$ give the same observational predictions than a quadratic potential and there is no need to extend the range to values much smaller than the Planck mass.  We chose to probe $\log_{10} \mu $ (from now $\mu$ and $\phic$ are given in reduced Planck mass units to lighten the notation) within the range $(-2 , 1.2)$, so that the entire transitory regime is probed.    

When $\phic$ is much larger than the Planck mass, the spectral index is too low for being acceptable, and thus the posterior probability is expected to be negligible.   The considered range for $\log_{10} \phic  $is $(-8,2) $, the upper bound being fixed so that all the region compatible with observations is included.   The lower bound is arbitrary.  We have checked that below this value the observable predictions are independent of $\phic$.   

For the reheating parameter $\ln R$, we have followed Ref.~\cite{Martin:2013nzq} and consider the range (-46,15), fixed so that it encompasses all the possible reheating histories with a reheating energy that cannot be lower than the BBN scale.  For $\ln As$ and other cosmological parameters ($\Omega_b h^2$, $\Omega_c h^2$, $\tau$, $\theta$) we have considered the default bounds in $\COSMOMC$, as well as for the 14 nuisance parameters of Planck.

\begin{figure*}[htb]
  \begin{center}
 \hspace{-1cm} \resizebox{0.8\textwidth}{!}{\includegraphics[angle=0,origin=c]{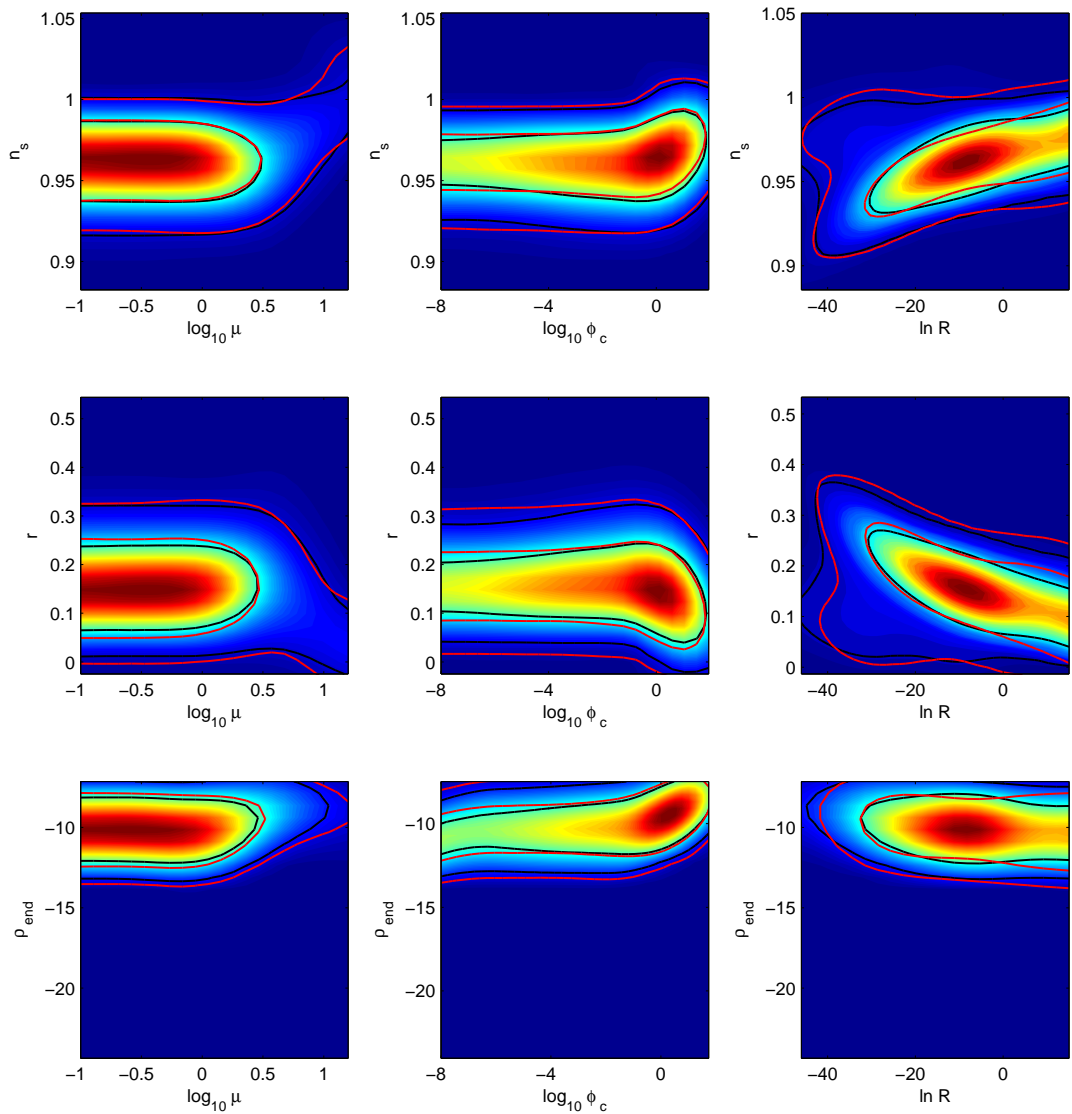}}
  \end{center}
  \caption{ Two-dimensional marginalized posterior probabilities for the hybrid model parameters (in reduced Planck mass units) as well as the derived parameters $n_s$, $r$ and $\rho_{\rr{end}}$ for Planck+BICEP2.  The red contours are the $1\sigma$ and $2\sigma$ regions of confidence.  The black contours are the $1\sigma$ and $2\sigma$ regions for Planck only.  }
  \label{fig:MCMC2D}
\end{figure*}

\subsection{Sampling method}

We have modified \COSMOMC so that the primordial power spectra are calculated using our external code for integrating the exact homogeneous dynamics and derive the spectral index and tensor to scalar ratio.   We thus include in $\COSMOMC$ the additional parameters $\log_{10} \phic $, $\log_{10} \mu$ and $\ln R$, as well as the derived parameters $n_{\rr s}, r, \rho_{\rr{end}}$ and $\Lambda$.

The parameter space to probe is therefore 22nd dimensional.  Posterior probabilities of hybrid model parameters are marginalized over the 18 other cosmological and nuisance parameters.   Because of the high-dimensionality of the parameter space, the sampling method must be such that the Markov chains must converge rapidly to optimize the computational time cost.  Therefore our external code is called only when at least one of the four model parameters is changed.  This is of importance when the fast-slow and fast dragging options are used instead of a simple vanilla Metropolis-Hastings algorithm.   This efficient sampling method has been proposed and described in Ref.~\cite{Lewis:2013hha}.  In a few words, two types of parameters (fast and slow) are considered, depending on how much it is computationally expansive to derived the likelihood when they are changed.  In addition, a fast dragging method is implemented, which decorrelate some parameters by rotating the sampling directions.  All together, this makes an important reduction of the computational cost compared to a standard Metropolis-Hastings algorithm.   Finally, note that the MCMC temperature has to be adjusted in order to optimize the sampling rate of the fast and slow parameter spaces.

\subsection{Statistical analysis results}

The Bayesian analysis has been conducted for Planck+BAO+BICEP2 data, as well as for Planck+BAO only.  
The one-dimensional and two-dimensional marginalized posterior probability density distributions for our model parameters $\log_{10} \mu$, $\log_{10} \phic$, $\ln A_s$ and $\ln R$ are displayed on Fig.~\ref{fig:MCMCtri}.   Posterior probabilities for the standard cosmological parameters are identical to the Planck analysis of a $\Lambda$-CDM model with $n_s$, $r$ and $\ln A_s$ as primordial spectra parameters.   This is expected given that our code derives $\ns$ and $r$ for each set of hybrid model parameters.   The marginalized probabilities for the derived parameters $r$, $n_s$ and $\rho_{\rr{end}} $ are displayed on Fig.~\ref{fig:MCMC2D}

We find that marginalized probabilities in the plane ($\log_{10} \mu$, $\log_{10} \phic$) are consistent with what was expected from Fig.~\ref{fig:contourns} and Fig.~\ref{fig:contourr}.   The likelihood is higher in the region corresponding to the transitory regime, and the best fit values given in Tab.~\ref{tab:bestfit} are located in this region.   
For Planck+BICEP2+BAO data, the best fit corresponds to a $\Delta \chi^2 \simeq 5.1$  in favor of hybrid inflation compared to the quadratic potential, with parameter values $\mu = 0.54 \Mpl$, $\phic = 6.4 \times 10^{-4} \Mpl$, $\ln R = -5 $.   Note however that the likelihood is reasonably flat in a rather wide region of the parameter space, which makes difficult to identify the best fit value.

The $1\sigma $ and $ 2\sigma$ intervals are reported in Tab.~\ref{tab:bestfit}.  For Planck+BICEP2+BAO we find that $\log_{10} \mu < 0.72 $ at $2\sigma$ level.  Above this value, the hybrid model corresponds to the usual picture of inflation in the vacuum dominated phase with a blue spectrum.   There is no lower bound on $\phic$.  Interestingly when BICEP2 data are included, the reheating parameter is constrained to be $ \ln R > -34 $.   At the same time, the energy density at the end of inflation has a maximum likelihood at $\rho_{\rr{end}} \sim 6 \times 10^{15} \rr{GeV}$, close to the GUT scale.   The energy scale of inflation lies within the range $ 4.3 \times 10^{15} \rr{GeV}< \rho_{\rr{end}}^{1/4} < 1.8 \times 10^{16} \rr{GeV}$ at $2 \sigma$ level.  

The chaotic regime remains within the $1\sigma$ bound, whereas for the large critical field regime we find that $\log_{10} \phic < 1.5$ at $2\sigma$ confidence level.  This bound is larger than what is expected from Fig.~\ref{fig:largephic} with $N_* = 60$, but it is obtained after marginalization over $\ln R$, which allows $N_*$ (and corresponding $\phi_*$) to take lower values.  

\begin{figure}[htb]
  \begin{center}
    \resizebox{0.5\textwidth}{!}{\includegraphics[angle=0,origin=c]{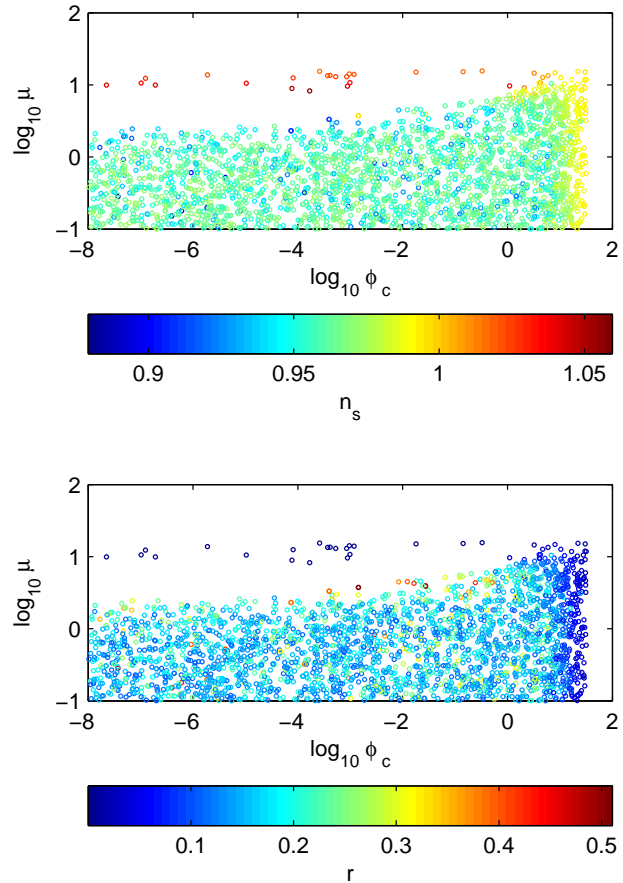}}
  \end{center}
  \vspace{-1cm}
  \caption{ Distribution of 3000 points within the Markov chains in the plane ($\log_{10} \mu, \log_{10} \phic$).  In the upper panel, the color scale represents the corresponding spectral index value, in the lower panel it represents the corresponding tensor to scalar ratio.  }
  \label{fig:MCMC3D}
\end{figure}

Finally we have displayed in Fig.~\ref{fig:MCMC3D}  the spectral index and the tensor to scalar ratio in the plane ($\log_{10} \mu, \log_{10} \phic$) for 3000 points of the Markov chains.   This figure illustrates that in the large critical field regime the spectral index is enhanced whereas the tensor to scalar ratio decreases.  It show also a few points at larger values of $\mu$ corresponding to the small field regime that generates a spectral index larger than unity.

Those results therefore confirm the analysis of the previous section for a fixed value of $N_*$ and give new constraints on the model parameters.   

\begin{table}[ht!]
\begin{center}
\begin{tabular}{|c|c|c|c|c|}
\hline
 Parameter & Best-fit & Mean &  $1\sigma$ range & $2\sigma$ range  \\
\hline
$\log_{10} \mu$ & -0.27  &  -0.18   & [ *,  0.045]         &  [*, *]          \\
$\log_{10} \phic $ &  -2.9 & -2.7     & [-4.6, 1.4 ]     & [*, 1.20 ]  \\
$\ln R $ & -8.4 & -8.0  & [-17, *] & [-32,*]  \\
$ \log_{10} \rho_{\rr{end}} $ & -10.5  & -10 & [-11, -9.5]& [-11, -8.3] \\
$n_s$ &  0.962 & 0.963 & [0.949, 0.977] & [0.929, 0.995] \\
$ r $ &  0.151   & 0.155 & [0.09, 0.20] & [0.03, 0.30] \\
\hline
$\log_{10} \mu$ & -0.24  &  -0.21   & [ *,  0.021]         &  [*, 0.72]          \\
$\log_{10} \phic $ &  -3.2  & -2.5     & [-4.0, 1.5 ]     & [*, 1.5 ]  \\
$\ln R $ & -5.00 & -10  & [-17, *] & [-34, *]  \\
$ \log_{10} \rho_{\rr{end}} $ & -10.5 & -10 & [-11, -8.9]& [-11, -8.4] \\
$n_s$ &  0.965 & 0.962 & [0.950, 0.975] & [0.930, 0.991] \\
$ r $ &  0.139   & 0.158 & [0.10, 0.20] & [0.04, 0.29] \\
\hline
\end{tabular} 
\end{center}

\caption{\label{tab:bestfit}  Best fit, mean likelihood and $1\sigma$ and $2\sigma$ intervals for hybrid model parameters (in units of reduced Planck mass), for Planck+BAO (upper part) and Planck+BAO+BICEP2 (lower part).  An asterisk denotes bounds not better than the prior limits.   } 
\end{table}


\section{Conclusion} \label{sec:ccl}

In the light of experimental results from Planck and BICEP2 experiments, we have re-analyzed the status of the original hybrid model for values of the field above the Planck scale.   Compared to previous analyses~\cite{Martin:2013nzq,Martin:2014lra}, we have included the effect of slow-roll violations between the large field and the vacuum dominated phases.   Using the exact background dynamics, we have identified three regimes of interests, with different observable predictions.  Then we have performed a Bayesian statistical analysis of the model parameter space and derived new constraints on the parameters.

A first regime of interest is the \emph{Large critical instability point } regime ($\phi_c > \mu $ and $ \phic \gtrsim M_\text{pl} $).  Inflation ends at super-Planckian field values with a nearly instantaneous tachyonic instability, where the potential in the direction of the inflationary valley is dominated by the quadratic term.   The observable predictions differ from the massive single field model, with a spectral index closer to unity and a lower tensor to scalar ratio.  The regime remains nevertheless within the $2\sigma$ confidence level of Planck and Planck+BICEP2 at the condition that $\phic \lesssim 10 \Mpl$.

In the second \emph{Chaotic-like} regime ($\phic < \mu \ll \mu_{\tt th} \simeq1.6 M_{\tt pl}$), the slow-roll is violated at the end of the large field phase and the field gains sufficient kinetic energy to overpass the vacuum dominated phase without reaching back the slow-roll attractor.  This non-trivial effect implies that the last 60 e-folds of inflation are realized in the large field phase, where the potential is dominated by the quadratic term.  Therefore the observable predictions cannot be distinguished from the massive single field model. 

The best statistical agreement with experimental data is found in the third \emph{Transitory} regime ($\phic < \mu \lesssim \mu_{\tt th}$).   In this case, due to transient slow-roll violations, several e-folds (typical between one and ten) are produced in the vacuum dominated phase but observable modes still leave the Hubble radius at large field values.   Compared to a massive single field model, the spectral index takes lower values and the tensor to scalar ratio is enhanced.  The best-fit to Planck data is found to be $\mu \sim 0.5 \Mpl$. 
The statistical analysis predicts a $\Delta \chi^2 \simeq 5.0$  in favor of hybrid inflation for the combined analysis of Planck and BICEP2 and $\Delta \chi^2 \simeq 0.9$ for Planck only.

Our analysis therefore demonstrates that the original hybrid model can lead to a red spectrum in agreement with the most recent observations, whereas in the common picture assuming slow-roll many e-folds of inflation occur generically in the small field phase, leading to a blue spectrum.   It can be noticed that in the \emph{transitory} regime parameters predicting a spectral index close to Planck best fit generate simultaneously a tensor to scalar ratio close to the central value observed by BICEP2.   Finally, note that future experiments such as COrE~\cite{2011arXiv1102.2181T} or PIXIE~\cite{2011JCAP...07..025K} will have the ability to distinguish between the three regimes identified above.  The transitory regime could also lead to an observable level of CMB distortions~\cite{Clesse:2014pna}.

\section*{Acknowledgements}

The authors warmly thank  Christophe Ringeval for useful discussions and comments.  The work of SC is supported by the \textit{mandat de retour} program of the Belgian Science Policy (BELSPO).  J.R. is supported by an FRS/FNRS Research Fellowship.
This research used ressources of the \textit{plateforme technologique en calcul intensif (PTCI)} of the University of Namur, Belgium, for which we acknowledge the financial support of the F.R.S.- FNRS (convention No. 2.4617.07. and 2.5020.11)

\bibliography{biblio_hybridLF}

\begin{thebibliography}{67}
\expandafter\ifx\csname natexlab\endcsname\relax\def\natexlab#1{#1}\fi
\expandafter\ifx\csname bibnamefont\endcsname\relax
  \def\bibnamefont#1{#1}\fi
\expandafter\ifx\csname bibfnamefont\endcsname\relax
  \def\bibfnamefont#1{#1}\fi
\expandafter\ifx\csname citenamefont\endcsname\relax
  \def\citenamefont#1{#1}\fi
\expandafter\ifx\csname url\endcsname\relax
  \def\url#1{\texttt{#1}}\fi
\expandafter\ifx\csname urlprefix\endcsname\relax\def\urlprefix{URL }\fi
\providecommand{\bibinfo}[2]{#2}
\providecommand{\eprint}[2][]{\url{#2}}

\bibitem[{\citenamefont{Ade et~al.}(2013{\natexlab{a}})}]{Ade:2013lta}
\bibinfo{author}{\bibfnamefont{P.}~\bibnamefont{Ade}} \bibnamefont{et~al.}
  (\bibinfo{collaboration}{Planck Collaboration})
  (\bibinfo{year}{2013}{\natexlab{a}}), \eprint{1303.5076}.

\bibitem[{\citenamefont{Ade et~al.}(2013{\natexlab{b}})}]{Planck:2013kta}
\bibinfo{author}{\bibfnamefont{P.}~\bibnamefont{Ade}} \bibnamefont{et~al.}
  (\bibinfo{collaboration}{Planck collaboration})
  (\bibinfo{year}{2013}{\natexlab{b}}), \eprint{1303.5075}.

\bibitem[{\citenamefont{Das et~al.}(2013)\citenamefont{Das, Louis, Nolta,
  Addison, Battistelli et~al.}}]{Das:2013zf}
\bibinfo{author}{\bibfnamefont{S.}~\bibnamefont{Das}},
  \bibinfo{author}{\bibfnamefont{T.}~\bibnamefont{Louis}},
  \bibinfo{author}{\bibfnamefont{M.~R.} \bibnamefont{Nolta}},
  \bibinfo{author}{\bibfnamefont{G.~E.} \bibnamefont{Addison}},
  \bibinfo{author}{\bibfnamefont{E.~S.} \bibnamefont{Battistelli}},
  \bibnamefont{et~al.} (\bibinfo{year}{2013}), \eprint{1301.1037}.

\bibitem[{\citenamefont{Story et~al.}(2012)\citenamefont{Story, Reichardt, Hou,
  Keisler, Aird et~al.}}]{Story:2012wx}
\bibinfo{author}{\bibfnamefont{K.}~\bibnamefont{Story}},
  \bibinfo{author}{\bibfnamefont{C.}~\bibnamefont{Reichardt}},
  \bibinfo{author}{\bibfnamefont{Z.}~\bibnamefont{Hou}},
  \bibinfo{author}{\bibfnamefont{R.}~\bibnamefont{Keisler}},
  \bibinfo{author}{\bibfnamefont{K.}~\bibnamefont{Aird}}, \bibnamefont{et~al.}
  (\bibinfo{year}{2012}), \eprint{1210.7231}.

\bibitem[{\citenamefont{Martin et~al.}(2014{\natexlab{a}})\citenamefont{Martin,
  Ringeval, Trotta, and Vennin}}]{Martin:2013nzq}
\bibinfo{author}{\bibfnamefont{J.}~\bibnamefont{Martin}},
  \bibinfo{author}{\bibfnamefont{C.}~\bibnamefont{Ringeval}},
  \bibinfo{author}{\bibfnamefont{R.}~\bibnamefont{Trotta}}, \bibnamefont{and}
  \bibinfo{author}{\bibfnamefont{V.}~\bibnamefont{Vennin}},
  \bibinfo{journal}{JCAP} \textbf{\bibinfo{volume}{1403}}, \bibinfo{pages}{039}
  (\bibinfo{year}{2014}{\natexlab{a}}), \eprint{1312.3529}.

\bibitem[{\citenamefont{Ade et~al.}(2013{\natexlab{c}})}]{Ade:2013ydc}
\bibinfo{author}{\bibfnamefont{P.}~\bibnamefont{Ade}} \bibnamefont{et~al.}
  (\bibinfo{collaboration}{Planck Collaboration})
  (\bibinfo{year}{2013}{\natexlab{c}}), \eprint{1303.5084}.

\bibitem[{\citenamefont{Ade et~al.}(2014)}]{Ade:2014xna}
\bibinfo{author}{\bibfnamefont{P.}~\bibnamefont{Ade}} \bibnamefont{et~al.}
  (\bibinfo{collaboration}{BICEP2 Collaboration}) (\bibinfo{year}{2014}),
  \eprint{1403.3985}.

\bibitem[{\citenamefont{Mortonson and Seljak}(2014)}]{Mortonson:2014bja}
\bibinfo{author}{\bibfnamefont{M.~J.} \bibnamefont{Mortonson}}
  \bibnamefont{and} \bibinfo{author}{\bibfnamefont{U.}~\bibnamefont{Seljak}}
  (\bibinfo{year}{2014}), \eprint{1405.5857}.

\bibitem[{\citenamefont{Flauger et~al.}(2014)\citenamefont{Flauger, Hill, and
  Spergel}}]{Flauger:2014qra}
\bibinfo{author}{\bibfnamefont{R.}~\bibnamefont{Flauger}},
  \bibinfo{author}{\bibfnamefont{J.~C.} \bibnamefont{Hill}}, \bibnamefont{and}
  \bibinfo{author}{\bibfnamefont{D.~N.} \bibnamefont{Spergel}}
  (\bibinfo{year}{2014}), \eprint{1405.7351}.

\bibitem[{\citenamefont{Martin et~al.}(2013)\citenamefont{Martin, Ringeval, and
  Vennin}}]{Martin:2013tda}
\bibinfo{author}{\bibfnamefont{J.}~\bibnamefont{Martin}},
  \bibinfo{author}{\bibfnamefont{C.}~\bibnamefont{Ringeval}}, \bibnamefont{and}
  \bibinfo{author}{\bibfnamefont{V.}~\bibnamefont{Vennin}}
  (\bibinfo{year}{2013}), \eprint{1303.3787}.

\bibitem[{\citenamefont{Dvali et~al.}(1994)\citenamefont{Dvali, Shafi, and
  Schaefer}}]{Dvali:1994ms}
\bibinfo{author}{\bibfnamefont{G.~R.} \bibnamefont{Dvali}},
  \bibinfo{author}{\bibfnamefont{Q.}~\bibnamefont{Shafi}}, \bibnamefont{and}
  \bibinfo{author}{\bibfnamefont{R.~K.} \bibnamefont{Schaefer}},
  \bibinfo{journal}{Phys. Rev. Lett.} \textbf{\bibinfo{volume}{73}},
  \bibinfo{pages}{1886} (\bibinfo{year}{1994}), \eprint{hep-ph/9406319}.

\bibitem[{\citenamefont{Bin\'etruy and Dvali}(1996)}]{Binetruy:1996xj}
\bibinfo{author}{\bibfnamefont{P.}~\bibnamefont{Bin\'etruy}} \bibnamefont{and}
  \bibinfo{author}{\bibfnamefont{G.~R.} \bibnamefont{Dvali}},
  \bibinfo{journal}{Phys. Lett.} \textbf{\bibinfo{volume}{B388}},
  \bibinfo{pages}{241} (\bibinfo{year}{1996}), \eprint{hep-ph/9606342}.

\bibitem[{\citenamefont{Clauwens and Jeannerot}(2008)}]{Clauwens:2007wc}
\bibinfo{author}{\bibfnamefont{B.}~\bibnamefont{Clauwens}} \bibnamefont{and}
  \bibinfo{author}{\bibfnamefont{R.}~\bibnamefont{Jeannerot}},
  \bibinfo{journal}{JCAP} \textbf{\bibinfo{volume}{0803}}, \bibinfo{pages}{016}
  (\bibinfo{year}{2008}), \eprint{0709.2112}.

\bibitem[{\citenamefont{Lazarides and
  Panagiotakopoulos}(1995)}]{Lazarides:1995vr}
\bibinfo{author}{\bibfnamefont{G.}~\bibnamefont{Lazarides}} \bibnamefont{and}
  \bibinfo{author}{\bibfnamefont{C.}~\bibnamefont{Panagiotakopoulos}},
  \bibinfo{journal}{Phys. Rev.} \textbf{\bibinfo{volume}{D52}},
  \bibinfo{pages}{559} (\bibinfo{year}{1995}), \eprint{hep-ph/9506325}.

\bibitem[{\citenamefont{Kallosh and Linde}(2003)}]{Kallosh:2003ux}
\bibinfo{author}{\bibfnamefont{R.}~\bibnamefont{Kallosh}} \bibnamefont{and}
  \bibinfo{author}{\bibfnamefont{A.}~\bibnamefont{Linde}},
  \bibinfo{journal}{JCAP} \textbf{\bibinfo{volume}{0310}}, \bibinfo{pages}{008}
  (\bibinfo{year}{2003}), \eprint{hep-th/0306058}.

\bibitem[{\citenamefont{Jeannerot et~al.}(2002)\citenamefont{Jeannerot, Khalil,
  and Lazarides}}]{Jeannerot:2002wt}
\bibinfo{author}{\bibfnamefont{R.}~\bibnamefont{Jeannerot}},
  \bibinfo{author}{\bibfnamefont{S.}~\bibnamefont{Khalil}}, \bibnamefont{and}
  \bibinfo{author}{\bibfnamefont{G.}~\bibnamefont{Lazarides}},
  \bibinfo{journal}{JHEP} \textbf{\bibinfo{volume}{07}}, \bibinfo{pages}{069}
  (\bibinfo{year}{2002}), \eprint{hep-ph/0207244}.

\bibitem[{\citenamefont{Lazarides and Vamvasakis}(2007)}]{Lazarides:2007fh}
\bibinfo{author}{\bibfnamefont{G.}~\bibnamefont{Lazarides}} \bibnamefont{and}
  \bibinfo{author}{\bibfnamefont{A.}~\bibnamefont{Vamvasakis}},
  \bibinfo{journal}{Phys. Rev.} \textbf{\bibinfo{volume}{D76}},
  \bibinfo{pages}{083507} (\bibinfo{year}{2007}), \eprint{arXiv:0705.3786
  [hep-ph]}.

\bibitem[{\citenamefont{Halyo}(1996)}]{Halyo:1996pp}
\bibinfo{author}{\bibfnamefont{E.}~\bibnamefont{Halyo}},
  \bibinfo{journal}{Phys. Lett.} \textbf{\bibinfo{volume}{B387}},
  \bibinfo{pages}{43} (\bibinfo{year}{1996}), \eprint{hep-ph/9606423}.

\bibitem[{\citenamefont{Bin\'etruy et~al.}(2004)\citenamefont{Bin\'etruy,
  Dvali, Kallosh, and Van~Proeyen}}]{Binetruy:2004hh}
\bibinfo{author}{\bibfnamefont{P.}~\bibnamefont{Bin\'etruy}},
  \bibinfo{author}{\bibfnamefont{G.}~\bibnamefont{Dvali}},
  \bibinfo{author}{\bibfnamefont{R.}~\bibnamefont{Kallosh}}, \bibnamefont{and}
  \bibinfo{author}{\bibfnamefont{A.}~\bibnamefont{Van~Proeyen}},
  \bibinfo{journal}{Class. Quant. Grav.} \textbf{\bibinfo{volume}{21}},
  \bibinfo{pages}{3137} (\bibinfo{year}{2004}), \eprint{hep-th/0402046}.

\bibitem[{\citenamefont{Jeannerot et~al.}(2000)\citenamefont{Jeannerot, Khalil,
  Lazarides, and Shafi}}]{Jeannerot:2000sv}
\bibinfo{author}{\bibfnamefont{R.}~\bibnamefont{Jeannerot}},
  \bibinfo{author}{\bibfnamefont{S.}~\bibnamefont{Khalil}},
  \bibinfo{author}{\bibfnamefont{G.}~\bibnamefont{Lazarides}},
  \bibnamefont{and} \bibinfo{author}{\bibfnamefont{Q.}~\bibnamefont{Shafi}},
  \bibinfo{journal}{JHEP} \textbf{\bibinfo{volume}{10}}, \bibinfo{pages}{012}
  (\bibinfo{year}{2000}), \eprint{hep-ph/0002151}.

\bibitem[{\citenamefont{Jeannerot}(1997)}]{Jeannerot:1997is}
\bibinfo{author}{\bibfnamefont{R.}~\bibnamefont{Jeannerot}},
  \bibinfo{journal}{Phys. Rev.} \textbf{\bibinfo{volume}{D56}},
  \bibinfo{pages}{6205} (\bibinfo{year}{1997}), \eprint{hep-ph/9706391}.

\bibitem[{\citenamefont{Jeannerot et~al.}(2003)\citenamefont{Jeannerot, Rocher,
  and Sakellariadou}}]{Jeannerot:2003qv}
\bibinfo{author}{\bibfnamefont{R.}~\bibnamefont{Jeannerot}},
  \bibinfo{author}{\bibfnamefont{J.}~\bibnamefont{Rocher}}, \bibnamefont{and}
  \bibinfo{author}{\bibfnamefont{M.}~\bibnamefont{Sakellariadou}},
  \bibinfo{journal}{Phys. Rev.} \textbf{\bibinfo{volume}{D68}},
  \bibinfo{pages}{103514} (\bibinfo{year}{2003}), \eprint{hep-ph/0308134}.

\bibitem[{\citenamefont{Rocher and Sakellariadou}(2005)}]{Rocher:2004et}
\bibinfo{author}{\bibfnamefont{J.}~\bibnamefont{Rocher}} \bibnamefont{and}
  \bibinfo{author}{\bibfnamefont{M.}~\bibnamefont{Sakellariadou}},
  \bibinfo{journal}{JCAP} \textbf{\bibinfo{volume}{0503}}, \bibinfo{pages}{004}
  (\bibinfo{year}{2005}), \eprint{hep-ph/0406120}.

\bibitem[{\citenamefont{Fukuyama et~al.}(2008)\citenamefont{Fukuyama, Okada,
  and Osaka}}]{Fukuyama:2008dv}
\bibinfo{author}{\bibfnamefont{T.}~\bibnamefont{Fukuyama}},
  \bibinfo{author}{\bibfnamefont{N.}~\bibnamefont{Okada}}, \bibnamefont{and}
  \bibinfo{author}{\bibfnamefont{T.}~\bibnamefont{Osaka}},
  \bibinfo{journal}{JCAP} \textbf{\bibinfo{volume}{0809}}, \bibinfo{pages}{024}
  (\bibinfo{year}{2008}), \eprint{0806.4626}.

\bibitem[{\citenamefont{Fairbairn et~al.}(2003)\citenamefont{Fairbairn,
  Lopez~Honorez, and Tytgat}}]{Fairbairn:2003yx}
\bibinfo{author}{\bibfnamefont{M.}~\bibnamefont{Fairbairn}},
  \bibinfo{author}{\bibfnamefont{L.}~\bibnamefont{Lopez~Honorez}},
  \bibnamefont{and} \bibinfo{author}{\bibfnamefont{M.~H.~G.}
  \bibnamefont{Tytgat}}, \bibinfo{journal}{Phys. Rev.}
  \textbf{\bibinfo{volume}{D67}}, \bibinfo{pages}{101302}
  (\bibinfo{year}{2003}), \eprint{hep-ph/0302160}.

\bibitem[{\citenamefont{Dvali and Tye}(1999)}]{Dvali:1998pa}
\bibinfo{author}{\bibfnamefont{G.~R.} \bibnamefont{Dvali}} \bibnamefont{and}
  \bibinfo{author}{\bibfnamefont{S.~H.~H.} \bibnamefont{Tye}},
  \bibinfo{journal}{Phys. Lett.} \textbf{\bibinfo{volume}{B450}},
  \bibinfo{pages}{72} (\bibinfo{year}{1999}), \eprint{hep-ph/9812483}.

\bibitem[{\citenamefont{Koyama et~al.}(2004)\citenamefont{Koyama, Tachikawa,
  and Watari}}]{Koyama:2003yc}
\bibinfo{author}{\bibfnamefont{F.}~\bibnamefont{Koyama}},
  \bibinfo{author}{\bibfnamefont{Y.}~\bibnamefont{Tachikawa}},
  \bibnamefont{and} \bibinfo{author}{\bibfnamefont{T.}~\bibnamefont{Watari}},
  \bibinfo{journal}{Phys. Rev.} \textbf{\bibinfo{volume}{D69}},
  \bibinfo{pages}{106001} (\bibinfo{year}{2004}), \eprint{hep-th/0311191}.

\bibitem[{\citenamefont{Berkooz et~al.}(2005)\citenamefont{Berkooz, Dine, and
  Volansky}}]{Berkooz:2004yc}
\bibinfo{author}{\bibfnamefont{M.}~\bibnamefont{Berkooz}},
  \bibinfo{author}{\bibfnamefont{M.}~\bibnamefont{Dine}}, \bibnamefont{and}
  \bibinfo{author}{\bibfnamefont{T.}~\bibnamefont{Volansky}},
  \bibinfo{journal}{Phys. Rev.} \textbf{\bibinfo{volume}{D71}},
  \bibinfo{pages}{103502} (\bibinfo{year}{2005}), \eprint{hep-ph/0409226}.

\bibitem[{\citenamefont{Davis and Postma}(2008)}]{Davis:2008sa}
\bibinfo{author}{\bibfnamefont{S.~C.} \bibnamefont{Davis}} \bibnamefont{and}
  \bibinfo{author}{\bibfnamefont{M.}~\bibnamefont{Postma}},
  \bibinfo{journal}{JCAP} \textbf{\bibinfo{volume}{0804}}, \bibinfo{pages}{022}
  (\bibinfo{year}{2008}), \eprint{0801.2116}.

\bibitem[{\citenamefont{Brax et~al.}(2007)}]{Brax:2006yq}
\bibinfo{author}{\bibfnamefont{P.}~\bibnamefont{Brax}} \bibnamefont{et~al.},
  \bibinfo{journal}{JCAP} \textbf{\bibinfo{volume}{0701}}, \bibinfo{pages}{026}
  (\bibinfo{year}{2007}), \eprint{hep-th/0610195}.

\bibitem[{\citenamefont{Kachru et~al.}(2003)}]{Kachru:2003sx}
\bibinfo{author}{\bibfnamefont{S.}~\bibnamefont{Kachru}} \bibnamefont{et~al.},
  \bibinfo{journal}{JCAP} \textbf{\bibinfo{volume}{0310}}, \bibinfo{pages}{013}
  (\bibinfo{year}{2003}), \eprint{hep-ht/0308055}.

\bibitem[{\citenamefont{Felder et~al.}(2001{\natexlab{a}})}]{Felder:2000hj}
\bibinfo{author}{\bibfnamefont{G.~N.} \bibnamefont{Felder}}
  \bibnamefont{et~al.}, \bibinfo{journal}{Phys. Rev. Lett.}
  \textbf{\bibinfo{volume}{87}}, \bibinfo{pages}{011601}
  (\bibinfo{year}{2001}{\natexlab{a}}), \eprint{hep-ph/0012142}.

\bibitem[{\citenamefont{Felder et~al.}(2001{\natexlab{b}})\citenamefont{Felder,
  Kofman, and Linde}}]{Felder:2001kt}
\bibinfo{author}{\bibfnamefont{G.~N.} \bibnamefont{Felder}},
  \bibinfo{author}{\bibfnamefont{L.}~\bibnamefont{Kofman}}, \bibnamefont{and}
  \bibinfo{author}{\bibfnamefont{A.~D.} \bibnamefont{Linde}},
  \bibinfo{journal}{Phys. Rev.} \textbf{\bibinfo{volume}{D64}},
  \bibinfo{pages}{123517} (\bibinfo{year}{2001}{\natexlab{b}}),
  \eprint{hep-th/0106179}.

\bibitem[{\citenamefont{Copeland et~al.}(2002)\citenamefont{Copeland, Pascoli,
  and Rajantie}}]{Copeland:2002ku}
\bibinfo{author}{\bibfnamefont{E.~J.} \bibnamefont{Copeland}},
  \bibinfo{author}{\bibfnamefont{S.}~\bibnamefont{Pascoli}}, \bibnamefont{and}
  \bibinfo{author}{\bibfnamefont{A.}~\bibnamefont{Rajantie}},
  \bibinfo{journal}{Phys. Rev.} \textbf{\bibinfo{volume}{D65}},
  \bibinfo{pages}{103517} (\bibinfo{year}{2002}), \eprint{hep-ph/0202031}.

\bibitem[{\citenamefont{Clesse}(2011)}]{Clesse:2010iz}
\bibinfo{author}{\bibfnamefont{S.}~\bibnamefont{Clesse}},
  \bibinfo{journal}{Phys. Rev.} \textbf{\bibinfo{volume}{D83}},
  \bibinfo{pages}{063518} (\bibinfo{year}{2011}), \eprint{1006.4522}.

\bibitem[{\citenamefont{Clesse et~al.}(2013)\citenamefont{Clesse, Garbrecht,
  and Zhu}}]{Clesse:2013jra}
\bibinfo{author}{\bibfnamefont{S.}~\bibnamefont{Clesse}},
  \bibinfo{author}{\bibfnamefont{B.}~\bibnamefont{Garbrecht}},
  \bibnamefont{and} \bibinfo{author}{\bibfnamefont{Y.}~\bibnamefont{Zhu}}
  (\bibinfo{year}{2013}), \eprint{1304.7042}.

\bibitem[{\citenamefont{Clesse and Garbrecht}(2012)}]{Clesse:2012dw}
\bibinfo{author}{\bibfnamefont{S.}~\bibnamefont{Clesse}} \bibnamefont{and}
  \bibinfo{author}{\bibfnamefont{B.}~\bibnamefont{Garbrecht}},
  \bibinfo{journal}{Phys.Rev.} \textbf{\bibinfo{volume}{D86}},
  \bibinfo{pages}{023525} (\bibinfo{year}{2012}), \eprint{1204.3540}.

\bibitem[{\citenamefont{Abolhasani et~al.}(2010)\citenamefont{Abolhasani,
  Firouzjahi, and Namjoo}}]{Abolhasani:2010kn}
\bibinfo{author}{\bibfnamefont{A.~A.} \bibnamefont{Abolhasani}},
  \bibinfo{author}{\bibfnamefont{H.}~\bibnamefont{Firouzjahi}},
  \bibnamefont{and} \bibinfo{author}{\bibfnamefont{M.~H.} \bibnamefont{Namjoo}}
  (\bibinfo{year}{2010}), \eprint{1010.6292}.

\bibitem[{\citenamefont{Mulryne et~al.}(2011)\citenamefont{Mulryne, Orani, and
  Rajantie}}]{Mulryne:2011ni}
\bibinfo{author}{\bibfnamefont{D.}~\bibnamefont{Mulryne}},
  \bibinfo{author}{\bibfnamefont{S.}~\bibnamefont{Orani}}, \bibnamefont{and}
  \bibinfo{author}{\bibfnamefont{A.}~\bibnamefont{Rajantie}},
  \bibinfo{journal}{Phys. Rev.} \textbf{\bibinfo{volume}{D84}},
  \bibinfo{pages}{123527} (\bibinfo{year}{2011}), \eprint{1107.4739}.

\bibitem[{\citenamefont{Kodama et~al.}(2011)\citenamefont{Kodama, Kohri, and
  Nakayama}}]{Kodama:2011vs}
\bibinfo{author}{\bibfnamefont{H.}~\bibnamefont{Kodama}},
  \bibinfo{author}{\bibfnamefont{K.}~\bibnamefont{Kohri}}, \bibnamefont{and}
  \bibinfo{author}{\bibfnamefont{K.}~\bibnamefont{Nakayama}},
  \bibinfo{journal}{Prog. Theor. Phys.} \textbf{\bibinfo{volume}{126}},
  \bibinfo{pages}{331} (\bibinfo{year}{2011}), \eprint{1102.5612}.

\bibitem[{\citenamefont{Linde}(1994)}]{Linde:1993cn}
\bibinfo{author}{\bibfnamefont{A.~D.} \bibnamefont{Linde}},
  \bibinfo{journal}{Phys. Rev.} \textbf{\bibinfo{volume}{D49}},
  \bibinfo{pages}{748} (\bibinfo{year}{1994}), \eprint{astro-ph/9307002}.

\bibitem[{\citenamefont{Copeland et~al.}(1994)\citenamefont{Copeland, Liddle,
  Lyth, Stewart, and Wands}}]{Copeland:1994vg}
\bibinfo{author}{\bibfnamefont{E.~J.} \bibnamefont{Copeland}},
  \bibinfo{author}{\bibfnamefont{A.~R.} \bibnamefont{Liddle}},
  \bibinfo{author}{\bibfnamefont{D.~H.} \bibnamefont{Lyth}},
  \bibinfo{author}{\bibfnamefont{E.~D.} \bibnamefont{Stewart}},
  \bibnamefont{and} \bibinfo{author}{\bibfnamefont{D.}~\bibnamefont{Wands}},
  \bibinfo{journal}{Phys. Rev.} \textbf{\bibinfo{volume}{D49}},
  \bibinfo{pages}{6410} (\bibinfo{year}{1994}), \eprint{astro-ph/9401011}.

\bibitem[{\citenamefont{Garbrecht et~al.}(2006)\citenamefont{Garbrecht, Pallis,
  and Pilaftsis}}]{Garbrecht:2006az}
\bibinfo{author}{\bibfnamefont{B.}~\bibnamefont{Garbrecht}},
  \bibinfo{author}{\bibfnamefont{C.}~\bibnamefont{Pallis}}, \bibnamefont{and}
  \bibinfo{author}{\bibfnamefont{A.}~\bibnamefont{Pilaftsis}},
  \bibinfo{journal}{JHEP} \textbf{\bibinfo{volume}{12}}, \bibinfo{pages}{038}
  (\bibinfo{year}{2006}), \eprint{hep-ph/0605264}.

\bibitem[{\citenamefont{Battye et~al.}(2010)\citenamefont{Battye, Garbrecht,
  and Moss}}]{Battye:2010hg}
\bibinfo{author}{\bibfnamefont{R.}~\bibnamefont{Battye}},
  \bibinfo{author}{\bibfnamefont{B.}~\bibnamefont{Garbrecht}},
  \bibnamefont{and} \bibinfo{author}{\bibfnamefont{A.}~\bibnamefont{Moss}},
  \bibinfo{journal}{Phys. Rev.} \textbf{\bibinfo{volume}{D81}},
  \bibinfo{pages}{123512} (\bibinfo{year}{2010}), \eprint{1001.0769}.

\bibitem[{\citenamefont{Battye et~al.}(2006)\citenamefont{Battye, Garbrecht,
  and Moss}}]{Battye:2006pk}
\bibinfo{author}{\bibfnamefont{R.~A.} \bibnamefont{Battye}},
  \bibinfo{author}{\bibfnamefont{B.}~\bibnamefont{Garbrecht}},
  \bibnamefont{and} \bibinfo{author}{\bibfnamefont{A.}~\bibnamefont{Moss}},
  \bibinfo{journal}{JCAP} \textbf{\bibinfo{volume}{0609}}, \bibinfo{pages}{007}
  (\bibinfo{year}{2006}), \eprint{astro-ph/0607339}.

\bibitem[{\citenamefont{Pallis and Shafi}(2013)}]{Pallis:2013dxa}
\bibinfo{author}{\bibfnamefont{C.}~\bibnamefont{Pallis}} \bibnamefont{and}
  \bibinfo{author}{\bibfnamefont{Q.}~\bibnamefont{Shafi}},
  \bibinfo{journal}{Phys.Lett.} \textbf{\bibinfo{volume}{B725}},
  \bibinfo{pages}{327} (\bibinfo{year}{2013}), \eprint{1304.5202}.

\bibitem[{\citenamefont{Chialva and Mazumdar}(2014)}]{Chialva:2014rla}
\bibinfo{author}{\bibfnamefont{D.}~\bibnamefont{Chialva}} \bibnamefont{and}
  \bibinfo{author}{\bibfnamefont{A.}~\bibnamefont{Mazumdar}}
  (\bibinfo{year}{2014}), \eprint{1405.0513}.

\bibitem[{\citenamefont{Linde}(1990)}]{Linde:2005ht}
\bibinfo{author}{\bibfnamefont{A.~D.} \bibnamefont{Linde}},
  \bibinfo{journal}{Contemp.Concepts Phys.} \textbf{\bibinfo{volume}{5}},
  \bibinfo{pages}{1} (\bibinfo{year}{1990}), \eprint{hep-th/0503203}.

\bibitem[{\citenamefont{Arkani-Hamed
  et~al.}(2003{\natexlab{a}})\citenamefont{Arkani-Hamed, Cheng, Creminelli, and
  Randall}}]{ArkaniHamed:2003mz}
\bibinfo{author}{\bibfnamefont{N.}~\bibnamefont{Arkani-Hamed}},
  \bibinfo{author}{\bibfnamefont{H.-C.} \bibnamefont{Cheng}},
  \bibinfo{author}{\bibfnamefont{P.}~\bibnamefont{Creminelli}},
  \bibnamefont{and} \bibinfo{author}{\bibfnamefont{L.}~\bibnamefont{Randall}},
  \bibinfo{journal}{JCAP} \textbf{\bibinfo{volume}{0307}}, \bibinfo{pages}{003}
  (\bibinfo{year}{2003}{\natexlab{a}}), \eprint{hep-th/0302034}.

\bibitem[{\citenamefont{Arkani-Hamed
  et~al.}(2003{\natexlab{b}})\citenamefont{Arkani-Hamed, Cheng, Creminelli, and
  Randall}}]{ArkaniHamed:2003wu}
\bibinfo{author}{\bibfnamefont{N.}~\bibnamefont{Arkani-Hamed}},
  \bibinfo{author}{\bibfnamefont{H.-C.} \bibnamefont{Cheng}},
  \bibinfo{author}{\bibfnamefont{P.}~\bibnamefont{Creminelli}},
  \bibnamefont{and} \bibinfo{author}{\bibfnamefont{L.}~\bibnamefont{Randall}},
  \bibinfo{journal}{Phys.Rev.Lett.} \textbf{\bibinfo{volume}{90}},
  \bibinfo{pages}{221302} (\bibinfo{year}{2003}{\natexlab{b}}),
  \eprint{hep-th/0301218}.

\bibitem[{\citenamefont{Kaplan and Weiner}(2004)}]{Kaplan:2003aj}
\bibinfo{author}{\bibfnamefont{D.~E.} \bibnamefont{Kaplan}} \bibnamefont{and}
  \bibinfo{author}{\bibfnamefont{N.~J.} \bibnamefont{Weiner}},
  \bibinfo{journal}{JCAP} \textbf{\bibinfo{volume}{0402}}, \bibinfo{pages}{005}
  (\bibinfo{year}{2004}), \eprint{hep-ph/0302014}.

\bibitem[{\citenamefont{Clesse and Rocher}(2009)}]{Clesse:2008pf}
\bibinfo{author}{\bibfnamefont{S.}~\bibnamefont{Clesse}} \bibnamefont{and}
  \bibinfo{author}{\bibfnamefont{J.}~\bibnamefont{Rocher}},
  \bibinfo{journal}{Phys.Rev.} \textbf{\bibinfo{volume}{D79}},
  \bibinfo{pages}{103507} (\bibinfo{year}{2009}), \eprint{0809.4355}.

\bibitem[{\citenamefont{Kobayashi and Seto}(2014)}]{Kobayashi:2014rla}
\bibinfo{author}{\bibfnamefont{T.}~\bibnamefont{Kobayashi}} \bibnamefont{and}
  \bibinfo{author}{\bibfnamefont{O.}~\bibnamefont{Seto}}
  (\bibinfo{year}{2014}), \eprint{1404.3102}.

\bibitem[{\citenamefont{Antusch et~al.}(2014)\citenamefont{Antusch, Cefala,
  Nolde, and Orani}}]{Antusch:2014saa}
\bibinfo{author}{\bibfnamefont{S.}~\bibnamefont{Antusch}},
  \bibinfo{author}{\bibfnamefont{F.}~\bibnamefont{Cefala}},
  \bibinfo{author}{\bibfnamefont{D.}~\bibnamefont{Nolde}}, \bibnamefont{and}
  \bibinfo{author}{\bibfnamefont{S.}~\bibnamefont{Orani}}
  (\bibinfo{year}{2014}), \eprint{1406.1424}.

\bibitem[{\citenamefont{Gong and Stewart}(2001)}]{Gong:2001he}
\bibinfo{author}{\bibfnamefont{J.-O.} \bibnamefont{Gong}} \bibnamefont{and}
  \bibinfo{author}{\bibfnamefont{E.~D.} \bibnamefont{Stewart}},
  \bibinfo{journal}{Phys.Lett.} \textbf{\bibinfo{volume}{B510}},
  \bibinfo{pages}{1} (\bibinfo{year}{2001}), \eprint{astro-ph/0101225}.

\bibitem[{\citenamefont{{Martin} and {Ringeval}}(2010)}]{2010PhRvD..82b3511M}
\bibinfo{author}{\bibfnamefont{J.}~\bibnamefont{{Martin}}} \bibnamefont{and}
  \bibinfo{author}{\bibfnamefont{C.}~\bibnamefont{{Ringeval}}},
  \bibinfo{journal}{\prd} \textbf{\bibinfo{volume}{82}}, \bibinfo{eid}{023511}
  (\bibinfo{year}{2010}), \eprint{1004.5525}.

\bibitem[{\citenamefont{Clesse et~al.}(2009)\citenamefont{Clesse, Ringeval, and
  Rocher}}]{Clesse:2009ur}
\bibinfo{author}{\bibfnamefont{S.}~\bibnamefont{Clesse}},
  \bibinfo{author}{\bibfnamefont{C.}~\bibnamefont{Ringeval}}, \bibnamefont{and}
  \bibinfo{author}{\bibfnamefont{J.}~\bibnamefont{Rocher}},
  \bibinfo{journal}{Phys. Rev.} \textbf{\bibinfo{volume}{D80}},
  \bibinfo{pages}{123534} (\bibinfo{year}{2009}), \eprint{0909.0402}.

\bibitem[{\citenamefont{Clesse}(2010)}]{Clesse:2009zd}
\bibinfo{author}{\bibfnamefont{S.}~\bibnamefont{Clesse}}, \bibinfo{journal}{AIP
  Conf. Proc.} \textbf{\bibinfo{volume}{1241}}, \bibinfo{pages}{543}
  (\bibinfo{year}{2010}), \eprint{0910.3819}.

\bibitem[{\citenamefont{Easther and Price}(2013)}]{Easther:2013bga}
\bibinfo{author}{\bibfnamefont{R.}~\bibnamefont{Easther}} \bibnamefont{and}
  \bibinfo{author}{\bibfnamefont{L.~C.} \bibnamefont{Price}}
  (\bibinfo{year}{2013}), \eprint{1304.4244}.

\bibitem[{\citenamefont{Easther et~al.}(2014)\citenamefont{Easther, Price, and
  Rasero}}]{Easther:2014zga}
\bibinfo{author}{\bibfnamefont{R.}~\bibnamefont{Easther}},
  \bibinfo{author}{\bibfnamefont{L.~C.} \bibnamefont{Price}}, \bibnamefont{and}
  \bibinfo{author}{\bibfnamefont{J.}~\bibnamefont{Rasero}}
  (\bibinfo{year}{2014}), \eprint{1406.2869}.

\bibitem[{\citenamefont{Lewis}(2013)}]{Lewis:2013hha}
\bibinfo{author}{\bibfnamefont{A.}~\bibnamefont{Lewis}},
  \bibinfo{journal}{Phys. Rev.} \textbf{\bibinfo{volume}{D87}},
  \bibinfo{pages}{103529} (\bibinfo{year}{2013}), \eprint{1304.4473}.

\bibitem[{\citenamefont{Martin and Vennin}(2012)}]{Martin:2011ib}
\bibinfo{author}{\bibfnamefont{J.}~\bibnamefont{Martin}} \bibnamefont{and}
  \bibinfo{author}{\bibfnamefont{V.}~\bibnamefont{Vennin}},
  \bibinfo{journal}{Phys.Rev.} \textbf{\bibinfo{volume}{D85}},
  \bibinfo{pages}{043525} (\bibinfo{year}{2012}), \eprint{1110.2070}.

\bibitem[{\citenamefont{Perreault~Levasseur
  et~al.}(2013)\citenamefont{Perreault~Levasseur, Vennin, and
  Brandenberger}}]{Levasseur:2013tja}
\bibinfo{author}{\bibfnamefont{L.}~\bibnamefont{Perreault~Levasseur}},
  \bibinfo{author}{\bibfnamefont{V.}~\bibnamefont{Vennin}}, \bibnamefont{and}
  \bibinfo{author}{\bibfnamefont{R.}~\bibnamefont{Brandenberger}},
  \bibinfo{journal}{Phys.Rev.} \textbf{\bibinfo{volume}{D88}},
  \bibinfo{pages}{083538} (\bibinfo{year}{2013}), \eprint{1307.2575}.

\bibitem[{\citenamefont{Martin et~al.}(2014{\natexlab{b}})\citenamefont{Martin,
  Ringeval, Trotta, and Vennin}}]{Martin:2014lra}
\bibinfo{author}{\bibfnamefont{J.}~\bibnamefont{Martin}},
  \bibinfo{author}{\bibfnamefont{C.}~\bibnamefont{Ringeval}},
  \bibinfo{author}{\bibfnamefont{R.}~\bibnamefont{Trotta}}, \bibnamefont{and}
  \bibinfo{author}{\bibfnamefont{V.}~\bibnamefont{Vennin}}
  (\bibinfo{year}{2014}{\natexlab{b}}), \eprint{1405.7272}.

\bibitem[{\citenamefont{{The COrE Collaboration}
  et~al.}(2011)\citenamefont{{The COrE Collaboration}, {Armitage-Caplan},
  {Avillez}, {Barbosa}, {Banday}, {Bartolo}, {Battye}, {Bernard}, {de
  Bernardis}, {Basak} et~al.}}]{2011arXiv1102.2181T}
\bibinfo{author}{\bibnamefont{{The COrE Collaboration}}},
  \bibinfo{author}{\bibfnamefont{C.}~\bibnamefont{{Armitage-Caplan}}},
  \bibinfo{author}{\bibfnamefont{M.}~\bibnamefont{{Avillez}}},
  \bibinfo{author}{\bibfnamefont{D.}~\bibnamefont{{Barbosa}}},
  \bibinfo{author}{\bibfnamefont{A.}~\bibnamefont{{Banday}}},
  \bibinfo{author}{\bibfnamefont{N.}~\bibnamefont{{Bartolo}}},
  \bibinfo{author}{\bibfnamefont{R.}~\bibnamefont{{Battye}}},
  \bibinfo{author}{\bibfnamefont{J.}~\bibnamefont{{Bernard}}},
  \bibinfo{author}{\bibfnamefont{P.}~\bibnamefont{{de Bernardis}}},
  \bibinfo{author}{\bibfnamefont{S.}~\bibnamefont{{Basak}}},
  \bibnamefont{et~al.}, \bibinfo{journal}{ArXiv e-prints}
  (\bibinfo{year}{2011}), \eprint{1102.2181}.

\bibitem[{\citenamefont{{Kogut} et~al.}(2011)\citenamefont{{Kogut}, {Fixsen},
  {Chuss}, {Dotson}, {Dwek}, {Halpern}, {Hinshaw}, {Meyer}, {Moseley},
  {Seiffert} et~al.}}]{2011JCAP...07..025K}
\bibinfo{author}{\bibfnamefont{A.}~\bibnamefont{{Kogut}}},
  \bibinfo{author}{\bibfnamefont{D.~J.} \bibnamefont{{Fixsen}}},
  \bibinfo{author}{\bibfnamefont{D.~T.} \bibnamefont{{Chuss}}},
  \bibinfo{author}{\bibfnamefont{J.}~\bibnamefont{{Dotson}}},
  \bibinfo{author}{\bibfnamefont{E.}~\bibnamefont{{Dwek}}},
  \bibinfo{author}{\bibfnamefont{M.}~\bibnamefont{{Halpern}}},
  \bibinfo{author}{\bibfnamefont{G.~F.} \bibnamefont{{Hinshaw}}},
  \bibinfo{author}{\bibfnamefont{S.~M.} \bibnamefont{{Meyer}}},
  \bibinfo{author}{\bibfnamefont{S.~H.} \bibnamefont{{Moseley}}},
  \bibinfo{author}{\bibfnamefont{M.~D.} \bibnamefont{{Seiffert}}},
  \bibnamefont{et~al.}, \bibinfo{journal}{JCAP} \textbf{\bibinfo{volume}{7}},
  \bibinfo{eid}{025} (\bibinfo{year}{2011}), \eprint{1105.2044}.

\bibitem[{\citenamefont{Clesse et~al.}(2014)\citenamefont{Clesse, Garbrecht,
  and Zhu}}]{Clesse:2014pna}
\bibinfo{author}{\bibfnamefont{S.}~\bibnamefont{Clesse}},
  \bibinfo{author}{\bibfnamefont{B.}~\bibnamefont{Garbrecht}},
  \bibnamefont{and} \bibinfo{author}{\bibfnamefont{Y.}~\bibnamefont{Zhu}}
  (\bibinfo{year}{2014}), \eprint{1402.2257}.

\end{thebibliography}

\end{document}